\documentclass[12pt,preprint]{aastex}
\shorttitle{Controlled fusion: inside stars}
\shortauthors{A. Ray}

\begin{document}

\title{Stars as thermonuclear reactors: their fuels and ashes\footnote
{Lecture notes on the 5th SERC School on
 Nuclear Physics, at Panjab
University, Chandigarh, Feb 11 - Mar 2, 2002, appearing in 
"Radioactive Ion Beams and Physics of Nuclei away
from the Line of Stability" (eds. Indra M. Govil and Rajiv K. Puri, 
Elite Publishing House, New Delhi 2003).
}
}

\author{A. Ray}
\affil{Tata Institute of Fundamental Research,
Mumbai 400 005, India; {\tt akr@tifr.res.in}}

%\def\rightmark{Controlled fusion: inside stars} 
%\def\leftmark{A. Ray}
 
%\title{Stars as thermonuclear reactors: their fuels and ashes} 
%\authors{
%
%{\twerm A. Ray$^{\; a}$ %
%}\\[2.812mm]
%{\normalsize
%\hspace*{-8pt}Tata Institute of Fundamental Research, 
%Mumbai 400 005, India\\[0.2ex] 
%
%}}

\begin{abstract} 
Atomic nuclei are transformed into each other in the cosmos by
nuclear reactions inside stars: -- the process of nucleosynthesis.
The basic concepts of determining nuclear reaction rates 
inside stars  and how they manage to burn their fuel so slowly most
of the time are discussed. Thermonuclear reactions 
involving protons in the hydrostatic burning of hydrogen in stars are
discussed first. 
This is followed by triple alpha reactions in the helium burning stage
and the issues of survival of carbon and oxygen in red giant stars connected
with nuclear structure of
oxygen and neon. Advanced stages of nuclear burning
in quiescent reactions involving carbon, neon, oxygen and silicon
are discussed. The role of neutron induced reactions in
nucleosynthesis beyond iron is discussed briefly, as also the 
experimental detection of neutrinos from SN1987A
which confirmed broadly the ideas concerning
gravitational collapse leading to a supernova. 

\end{abstract}

 
%\maketitle
%\setcounter{page}{1}
 
\section{Introduction}
Most people do not think of the sun as a star and few would consider
stars as nuclear reactors. Yet, not only that is the way it is, even our
own world is made out of the ``fall-out" 
from stars that blew up and spewed radioactive debris into
the nascent solar system.

Nuclear Astrophysics is the field concerning ``the synthesis and
Evolution of atomic nuclei, by thermonuclear reactions, from
the Big Bang to the present \cite{Arn96}. What is the origin of the 
matter of which we are made?".
Our high entropy universe, presumably resulting from the Big Bang,
contains many more photons per particle of matter with mass,
e.g. electrons, protons and neutrons. Because of the high entropy
and the consequent low density of matter (on terrestrial or
stellar scales) at any given temperature as the universe expanded,
there was time to manufacture elements 
only upto helium and the major products of
cosmic nucleosynthesis remained hydrogen and helium\footnote
{Note however suggestions (\cite{Car84}, \cite{Car94})
that early generation of stars called Pop III objects
can also contribute to the abundance of $^4He$ seen in the universe today
and therefore the entire present helium may not be a product of big
bang nucleosynthesis only, -- further aggravating the problems of theoretical
predictions of standard big bang nucleosynthesis compared to
observed abundances (\cite{Sal03}).}.
Stars formed
from this primordial matter and they used these elements as fuel
to generate energy like a giant nuclear reactor. In the process,
the stars could shine and manufacture 
higher atomic number elements like carbon, oxygen, calcium and iron
of which we and our world is made. The heavy elements are either dredged up
from the core of the star to the surface of the star from which they
are dispersed by stellar wind or directly ejected into the interstellar
medium when a (massive) star explodes. The stardust is the source of
heavy elements for new generation of stars and sun-like systems.
  
The sun is slowly burning a light element, namely, hydrogen into a heavier
element, helium. It is not exactly the same now as it just started burning
hydrogen in its core and will start to look noticeably different once
it exhausts all the hydrogen it {\it can} burn in its core. In other
words, nuclear reactions in the interiors of stars determine the evolution
or the life-cycle of the stars, apart from providing them with internal
power for heat and light and manufacturing all the heavier elements
that the early universe could not.

Since nuclear astrophysics is not usually taught at the 
master's level nuclear physics specialisation 
in our universities,
these lecture notes are meant to be an introduction to the subject and a
pointer to literature and internet resources.
(See for example, \cite{Hax99} for a course of nuclear astrophysics, and
the International Conference Proceedings under the title: ``Nuclei
in the Cosmos" for periodic research conferences in the field. 
Valuable nuclear astrophysics datasets in machine readable
forms useful for researchers can be found at sites:\cite{Ornl99}, \cite{Lbl98}.
Much of the material discussed here can be found in textbooks
in the subject, see e.g. \cite{Rol88}, \cite{Cla68}, \cite{Ree68}, 
\cite{Bah89}, \cite{Arn96} etc.). The
emphasis here is on the nuclear reactions in the stars and how these
are calculated, rather than how stars evolve. The latter usually form a core
area of stellar astrophysics. There is a correspondence between
the evolutionary state of a star, its external appearance and internal
core conditions and the nuclear fuel it burns, -- a sort of a mapping
between the astronomers Hertzsprung-Russel diagram and the
nuclear physicist's chart of the nuclides, until nuclear
burning takes place on too rapid a timescale -- see Reeves \cite{Ree68}.
This article is organised essentially
in the same sequence that a massive star burns successively
higher atomic number elements in its core, until it collapses
and explodes in a supernova. The introductory part 
discusses how the rates of thermonuclear reactions in stars are 
calculated, what the different classes of reactions are and how the
stars (usually) manage to burn their fuels so slowly.

\section{Stars and their thermonuclear reactions}

While referring to Sir Ernest Rutherford ``breaking down the atoms
of oxygen and nitrogen, driving out an isotope of helium from them",
Arthur Eddington remarked: ``what is possible in the Cavendish
Laboratory may not be too difficult in the sun" \cite{Edd20}.
Indeed this is the case, but for the fact that a star does
this by fusion reactions, rather than spallation reactions, --
in the process giving out heat and light and manufacturing fresh elements. 
Of all the light elements, hydrogen
is the most important one in this regard, because: a) it has a large
universal abundance, b) considerable energy evolution is possible
with it because of the large binding energies of nuclei that can
be generated from its burning and c) its small charge and mass
allows it to penetrate easily through the potential barriers of
other nuclei.
A long term goal of terrestrial plasma physicists has been to achieve a
sustained and controlled thermonuclear fusion at economical rates
in the laboratory. A star burns its fuel in the core quite naturally
via similar thermonuclear reactions, where the confinement of the fuel is
achieved in the star's own gravitational field. These reactions remain
``controlled", or self-regulated, as long as the stellar material
remains non-degenerate. (There are however examples to the contrary
when a whole white dwarf (resulting from an evolved intermediate mass star)
undergoes merger with another and explodes, as nuclear
fuel (carbon) is ignited under degenerate conditions, 
such as in a type Ia supernova). 

The recognition of the quantum mechanical tunneling effect prompted
Atkinson and Houtermans \cite{Atk29} to work out the qualitative treatment
of energy production in stars. They suggested (``how to cook a nucleus in a
pot") that the nucleus serves as both a cooking pot and a trap. 
Binding energy difference of four protons and two electrons 
(the nuclear fuel) and their ash, the helium nucleus, some 26.7 MeV
is converted to heat and light that we receive from the sun\footnote{
Lord Kelvin surmised in the nineteenth century that the solar
luminosity is supplied by the gravitational contraction of the sun. Given
the solar luminosity, this immediately defined a solar lifetime (the so-called
Kelvin-Helmholtz time): $\tau_{KH} = G M_{\odot}^2/R_{\odot}L_{\odot} \sim
\rm few \times 10^7 \rm yr$. This turned out to be much shorter
than the estimated age
of the earth at that time known from fossil records, and led to a famous debate
between Lord Kelvin and the fossil geologist Cuvier. In fact, as noted above
modern estimates of earth's age are much longer
and therefore the need to maintain
sunshine for such a long time requires that the amount of energy
radiated by the sun during its lifetime is much larger than its gravitational
energy or its internal (thermal) energy:
$L_{\odot} \times t_{life} \gg G M_{\odot}^2 / R_{\odot}$.
This puzzle was
resolved only later with the realisation that the star can tap its
much larger nuclear energy reservoir in its core through thermonuclear
reactions. The luminosity of the sun however is determined by an
interplay of atomic and gravitational physics that controls the opacity,
chemical composition, the balance of pressure forces against gravity, etc.
Nuclear physics determines how fast nuclear reactions should go under
the ambient conditions which regulate through feedback control
those reaction rates.
}. 
The photons in the interior are scattered many a times, for tens of
millions of years, before they reach the surface of the sun.
Some of the reactions also produce neutrinos, which because of their
small cross-section for interaction, are not stopped or scattered by
overlying matter, -- but just stream out straight from the core.
Neutrinos therefore are best probes of the core of the star (\cite{Dav68},
\cite{Hir87}), while the photons bear information from their surface of 
last scattering -- the photosphere of the star. 

\subsection{Why do the stars burn slowly: a look at Gamow peaks}

The sun has been burning for at least 4.6 billion years\footnote
{Lord Rutherford \cite{Rut29} determined the age of a sample
of pitchblende, to be 700 million years,
by measuring the amount of uranium and radium and helium retained in the rock
and by calculating the annual output of alpha particles. The oldest
rock found is from Southwest Greenland: $\approx 3.8$ Gyr old
\cite{Bah89}. Radioactive dating of meteorites point to their formation and the
solidification of the earth about $4.55 \pm 0.07$ years ago \cite{Kir78}.
Since the sun and the solar system formed only slightly before,
their age at isolation and condensation
from the interstellar medium is taken to be 4.6 Gyr \cite{Fow77}.
}.
How does it manage to burn so slowly\footnote
{The Nobel prize citation of Hans Bethe (1967) who solved this problem, noted:
" This year's Nobel Prize in Physics - to professor Hans A. Bethe
  - concerns an old riddle. How has it been possible for the sun
  to emit light and heat without exhausting its source not only
  during the thousands of centuries the human race has existed
  but also during the enormously long time when living beings
  needing the sun for their nourishment have developed and
  flourished on our earth thanks to this source? The solution of
  this problem seemed even more hopeless when better
  knowledge of the age of the earth was gained. None of the
  energy sources known of old could come under consideration.
  Some quite unknown process must be at work in the interior of
  the sun. Only when radioactivity, its energy generation
  exceeding by far any known fuel, was discovered, it began to
  look as if the riddle might be solved. And, although the first
  guess that the sun might contain a sufficient amount of
  radioactive substances soon proved to be wrong, the closer
  study of radioactivity would by and by open up a new field of
  physical research in which the solution was to be found.
......
  A very important part of his work resulted in eliminating a great
  number of thinkable nuclear processes under the conditions at
  the centre of the sun, after which only two possible processes
  remained..... (Bethe) attempted a thorough analysis of these and other
  thinkable processes necessary to make it reasonably certain
  that these processes, and only these, are responsible for the
  energy generation in the sun and similar stars.
"}?
Under the ambient conditions in the core,
the relevant thermonuclear reaction cross sections are very small\footnote
{This makes the experimental verification of the
reaction cross-sections a
very challenging task, requiring in some cases, extremely
high purity targets and projectiles
so that relevant small event rates are not swamped by other reaction channels
and products (see Rolfs and Rodney, Chapter 5 \cite{Rol88}).}.
For reactions involving charged particles, nuclear
physicists often encounter cross-sections near the Coulomb barrier
of the order of millibarns. One can obtain a characteristic luminosity
$L_C$ based on this cross section and 
the nuclear energy released per reaction \cite{Bah89} : 
$$ L_C \sim \epsilon N \Delta E / \tau_C$$
where $\epsilon \approx 10^{-2}$ is the fraction of total number
of solar nuclei $N \sim 10^{57}$ that take part in nuclear fusion reactions
generating typically $\Delta E \sim 25$ MeV in hydrogen to helium conversion.
Here, the $\tau_C$ is the characteristic timescale for reactions,
which becomes miniscule %($\sim 10^{-275}$)
for the cross-sections at the Coulomb barrier and the
ambient density and relative speed of the reactants etc:
$$ \tau_C \sim {1\over n \sigma v} = {10^{-8} s \over [n/(10^{26} \rm cm^{-3}]
[\sigma / 1 mbarn] [v /10^9 \rm cm s^{-1}] }$$
This would imply a characteristic luminosity of  
$L_c \approx 10^{20} L_{\odot}$, 
even for a small fraction of the solar material taking part in the
reactions (i.e. $\epsilon \sim 10^{-2}$). If this were
really the appropriate cross-section for the reaction,
the sun would have been gone very quickly indeed, 
Instead the cross-sections are much less than
that at the Coulomb barrier penetration energy (say at proton energies of
1 MeV), to allow for a long lifetime of the sun (in addition,
weak-interaction process gives a smaller cross-section for some reactions than
electromagnetic process, -- see Section 3.1). 

\begin{figure}[htb]
\vspace*{-1.0cm}
\hspace*{-3.0cm}
%                 \insertplot{fig1.eps}
\plotone{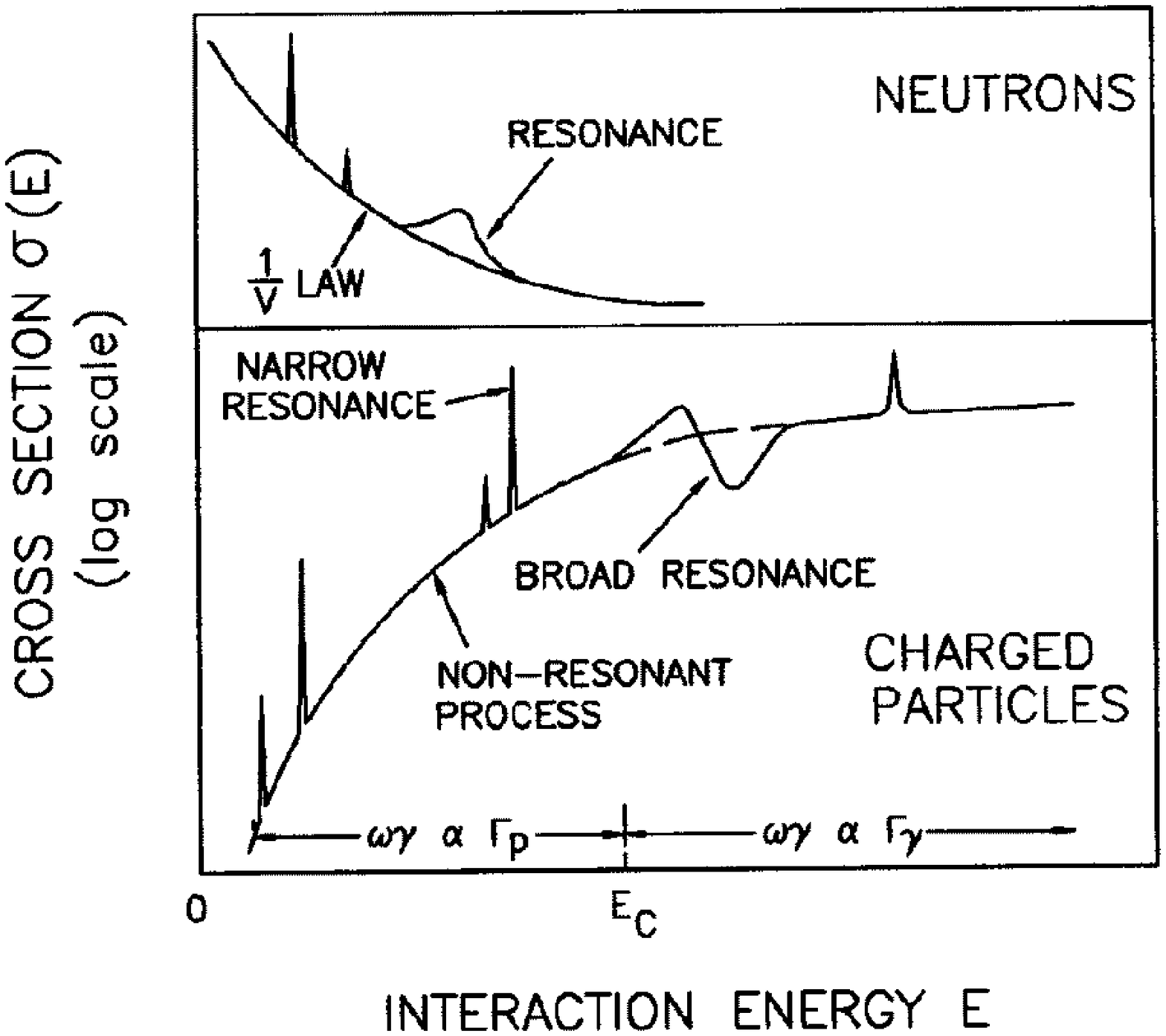}
\vspace*{-1.0cm}
\caption[]{
Dependence of total cross-sections on the interaction energy 
for neutrons (top panel) and charged particles (bottom panel). 
Note the presence
of resonances (narrow or broad) superimposed on a slowly varying
nonresonant cross-section (after \cite{Rol88}). 
%%\label{fig: cross-sections}
}
\label{fig: cross-sections}
\end{figure}

Stellar nuclear reactions can be either: a) charged particle reactions
(both target and projectile are nuclei) or b) neutral particle
(neutron) induced reactions. Both sets of reactions can go through
either a resonant state of an intermediate nucleus or can be a non-resonant
reaction. In the former reaction, the intermediate state could be a narrow
unstable state, which decays into other particles or nuclei. In general,
a given reaction can involve both types of reaction channels.
In charged particle induced reactions, the cross-section for both 
reaction mechanisms drops rapidly with decreasing energy, due to 
the effect of the Coulomb barrier (and thus it becomes more difficult 
to measure stellar reaction cross-sections accurately).
In contrast, the neutron induced reaction cross-section is very large and 
increases with decreasing energy (here, resonances may be superposed on a 
smooth non-resonant yield which follows the 
$1/v \sim 1/ \sqrt E$  dependence). 
These reaction rates and cross-sections can be then directly 
measured at stellar energies that are relevant 
(if such nuclei are long lived or can be generated).
The schematic dependence of the cross-sections are shown in Fig. 
\ref{fig: cross-sections}.

%\begin{figure}[htb]
%\vspace*{-1.0cm}
%\hspace*{-3.0cm}
%                 \insertplot{fig1.eps}
%\vspace*{-1.0cm}
%\caption[]{
%Dependence of total cross-sections on the interaction energy 
%for neutrons (top panel) and charged particles (bottom panel). 
%Note the presence
%of resonances (narrow or broad) superimposed on a slowly varying
%nonresonant cross-section (after \cite{Rol88}). 
%%\label{fig: cross-sections}
%}
%\label{fig: cross-sections}
%\end{figure}

\subsection{Gamow peak and the astrophysical S-factor}

The sun and other ``main-sequence" stars (burning hydrogen 
in their core quiescently) evolve very slowly by adjusting their central 
temperature such that the average  thermal energy of a nucleus is small 
compared to the Coulomb repulsion an ion-ion pair encounters. This is how
stars can live long for astronomically long times.
A central temperature $T \geq 10^7$K (or $T_7 \geq 1$, hereafter
a subscript x to temperature or density, indicates a temperature in
units of $10^x$) is required for sufficient
kinetic energy of the reactants to overcome the Coulomb barrier
and for thermonuclear reactions involving 
hydrogen to proceed at an effective rate, even though fusion reactions have 
positive Q  values i.e. net energy is liberated out of the reactions. 
The classical turning point radius for a projectile of charge $Z_2$ and
kinetic energy $E_p$ (in a Coulomb potential $V_C = Z_1 Z_2 e^2 / r $, and
effective height of the Coulomb barrier $E_C= Z_1 Z_2 e^2/ R_n = 550 \;\rm keV$
for a p + p reaction), is:
$r_{cl} = Z_1 Z_2 e^2 / E_p$.
Thus, classically a p + p reaction would proceed only when the kinetic energy
exceeds 550 keV. Since the number of particles traveling at a given speed
is given by the Maxwell Boltzmann (MB) distribution $\phi(E)$, only the tail
of the MB distribution above 550 keV is effective when
the typical thermal energy is 0.86 keV ( $T_9 = 0.01$). The ratio of
the tails of the MB distributions: 
$\phi(550 \; \rm keV) / \phi(0.86 \; \rm keV)$
is quite miniscule, and thus classically at typical stellar temperatures
this reaction will be virtually absent. 

\begin{figure}[htb]
\vspace*{-0.5cm}
%                 \insertplot{fig2.eps}
\plotone{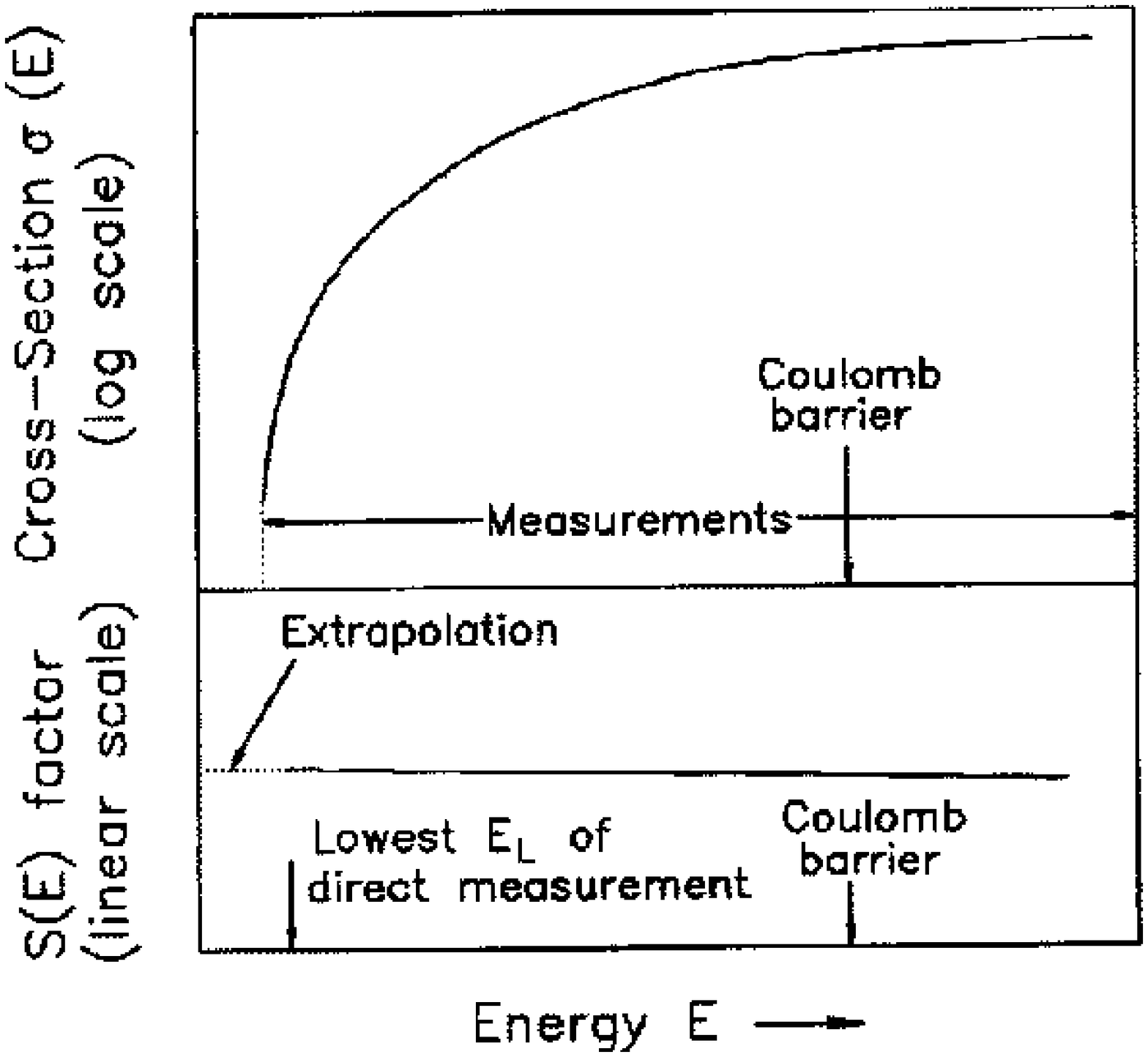}
\vspace*{-1.0cm}
\caption[]{
Cross-section and astrophysical S-factor for charged particle reactions
as a function of beam energy. The effective range of energy in stellar
interiors is usually far less than the Coulomb barrier energy $E_C$ or
the lower limit $E_L$ where laboratory measurements can be carried out.
Note that the scale is logarithmic for the cross-section but linear for
S-factor, and hence the cross section drops sharply in the region of
astrophysical interest, whereas the change is much less severe for the
S-factor. Therefore, necessary extrapolation of laboratory data to lower
energies relevant for astrophysical situations is more reliable in the
case of S-factor.
}
\label{fig: SigmavsS-fac}
\end{figure}

Although classically a particle
with projectile energy $E_p$ cannot penetrate beyond the classical turning
point, quantum mechanically, one has a finite value of the squared 
wave function at the nuclear radius $R_n : | \psi(R_n) |^2 $.
The probability that the incoming particle penetrates the barrier is:

$$ P = {| \psi(R_n) |^2 \over | \psi(R_c) |^2} $$
where $\psi(r)$ are the wavefunctions at corresponding points. Bethe 
\cite{Bet37} solved the Schroedinger equation for the Coulomb potential
and obtained the transmission probability:-

$$ P = \rm exp\bigg(-2 KR_c \big[{\rm tan^{-1}(R_c/R_n - 1)^{1/2}\over (R_c/R_n-1)^{1/2}}-{R_n\over R_c}\big]\bigg)$$
with $K=[2\mu / \hbar^2(E_c - E)]^{1/2}$.
This probability reduces to a much simpler relation
at the low energy limit: $E \ll E_c$, which is equivalent to the classical
turning point $R_c$ being much larger than the nuclear radius $R_n$.
The probability is:

$$P = \rm exp(-2\pi\eta) = exp[-2\pi Z_1 Z_2 e^2/(\hbar v)] = exp[-31.3Z_1Z_2({\mu \over E})^{1/2}]$$
where in the second equality, $\mu$ is the reduced mass in Atomic Mass Units
and E is the centre of mass energy in keV. The exponential
quantity involving the square brackets in the
second expression is called the ``Gamow factor". The reaction cross-section
between particles of charge $Z_1$ and $Z_2$ has this exponential dependence
due to the Gamow factor. In addition, because the cross-sections are essentially
``areas": proportional to $\pi(\lambda/2\pi\hbar)^2 \propto 1/E$, it is 
customary to write the cross-section, with these two energy dependences 
filtered out:

$$\sigma(E) = {\rm exp(-2\pi\eta)\over E} S(E)$$
where the factor $S(E)$ is called the astrophysical (or nuclear) S-factor.
The S-factor may contain degeneracy factors due to spin, e.g. 
$[(2J+1)/{(2J_1+1)(2J_2+1)}]$ as reaction cross-sections
are summed over final states and averaged over initial states.
Because the rapidly varying parts of the cross-section (with energy)
are thus filtered out, the S-factor is a slowly varying function
of center of mass energy, at least for the non-resonant reactions.
It is thus much safer to extrapolate $S(E)$ to the energies relevant
for astrophysical environments from the laboratory data, which is
usually generated at higher energies (due to difficulties of measuring
small cross-sections), than directly extrapolating the $\sigma(E)$,
which contains the Gamow transmission factor 
(see Fig. \ref{fig: SigmavsS-fac}).
Additionally, in order to relate $\sigma(E)$ and $S(E)$, quantities 
measured in the laboratory to these relevant quantities in the solar
interior, a correction factor $f_0$ due to the effects of electron screening
needs to be taken into account \cite{Sal54}.

\begin{figure}[htb]
\vspace*{-0.5cm}
%                 \insertplot{fig3.eps}
\plotone{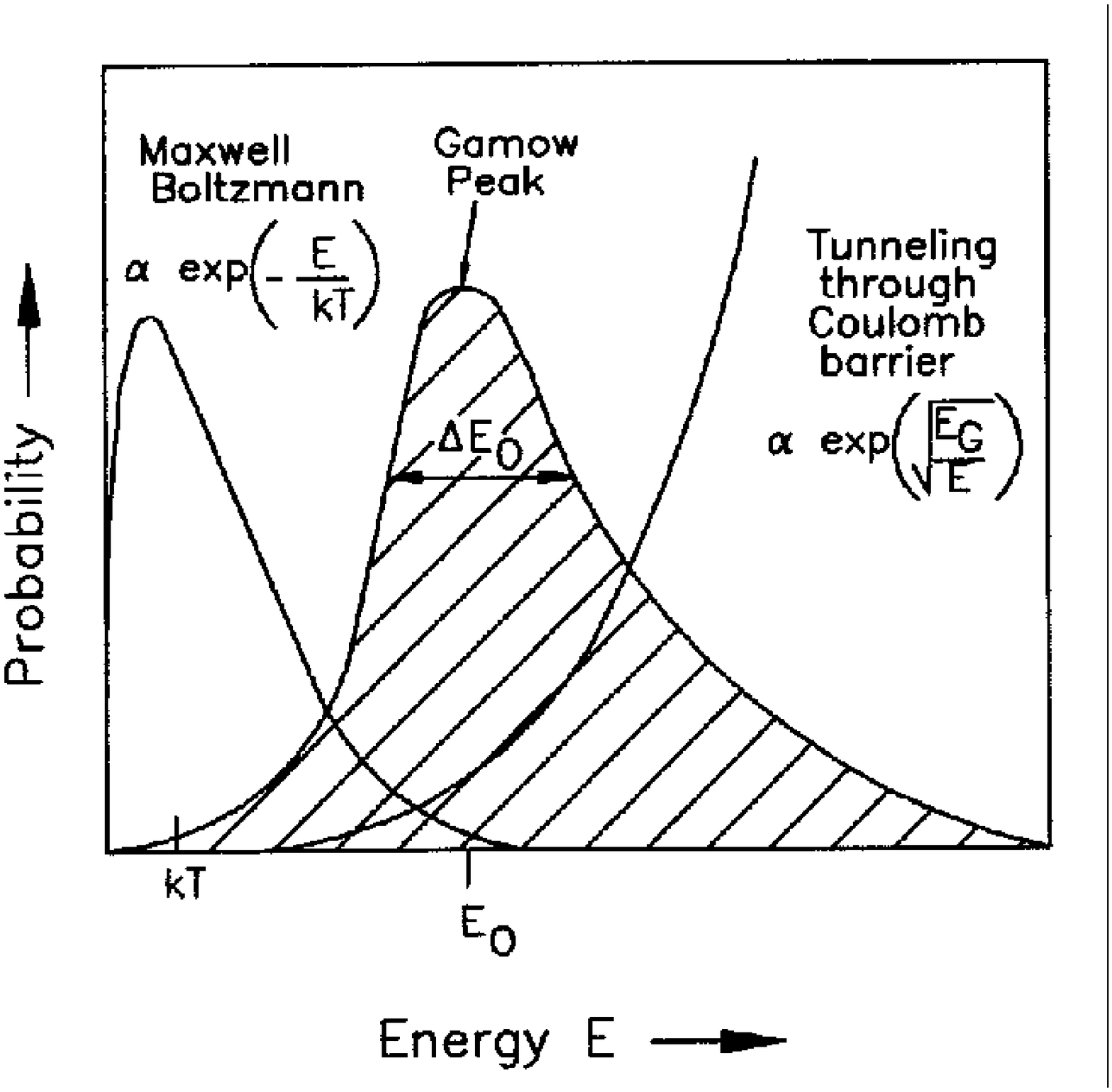}
\vspace*{-0.9cm}
\caption[]{
The Gamow peak is a convolution of the energy distribution of
the Maxwell Boltzmann probability and the quantum mechanical Coulomb 
barrier transmission probability. The peak in the shaded region near
energy $E_0$ is the Gamow peak that gives the highest probability 
for charged particle reactions to take place. Usually the Gamow peak is
at a much higher energy than $kT$, and in the figure the ordinate scale
(for the Gamow peak) is magnified with respect to those of the
M-B and barrier penetration factors. See also Table \ref{tab: parameters}.
}
\label{fig: gamowpeak}
\end{figure}

In the stellar core with a temperature T, reacting particles 
have many different velocities (energies) according to a Maxwell - 
Boltzmann distribution :-

$$ \phi(v) = 4\pi v^2 \bigg({\mu \over 2\pi kT}\bigg)^{3/2} \rm exp\bigg[-{\mu v^2 \over 2kT}\bigg]
 \propto E \; exp[-E/kT] $$
Nuclear cross-section or the reaction rates which also depend upon the
relative velocity (or equivalently the centre of mass energy) therefore need
to be averaged over the thermal velocity (energy) distribution. Therefore,
the thermally averaged reaction rate per particle pair is:

$$<\sigma v> = \int_0^{\infty} \phi(v) \sigma(v) v dv
             = \big({8\over \pi \mu}\big)^{1/2}{1\over(kT)^{3/2}} \int_0^{\infty} \sigma(E) E \; \rm exp(-E/kT) dE $$
The thermally averaged reaction rate per pair is, utilising the astrophysical
S-factor and the energy dependence of the Gamow-factor:

$$<\sigma v> = \big({8\over \pi\mu}\big)^{1/2}{1\over(kT)^{3/2}} \int_0^{\infty}\rm S(E) exp \big[ -{E\over kT} - {b\over \sqrt E} \big]dE $$
with $b^2 = E_G = 2\mu (\pi e^2 Z_1 Z_2/ \hbar)^2 = 0.978\mu Z_1^2Z_2^2$ MeV,
$E_G$ being called the Gamow energy.
Note that in the expression for the reaction rate above, at low energies,
the exponential term $\rm exp(-b/\sqrt E) = exp(-\sqrt(E_G/E))$ becomes
very small whereas at high energies the Maxwell-Boltzmann factor 
$\rm exp(-E/kT)$ vanishes. 
Hence there would be a peak (at energy, say, $E_0$)
of the integrand for the thermally averaged reaction rate per pair 
(see Fig. \ref{fig: gamowpeak}).
The exponential part of the energy integrand can be approximated as:

$$ \rm exp \big[-{E\over kT} - b E^{-1/2}\big] \sim C \; exp \bigg[-\big({E-E_0\over \Delta/2}\big)^2\bigg]$$
where 

$$C= \rm exp(-E_0/kT - bE_0^{-1/2}) = exp(-3E_0/kT) = exp(-\tau)$$ 
with
$$E_0 = (b kT/2)^{{2\over3}} = 1.22 \rm keV (Z_1^2 Z_2^2 \mu T_6^2)^{{1\over3}}$$ 
and

$$\Delta = 4(E_0 kT/3)^{{1\over2}} = 0.75 \rm keV (Z_1^2 Z_2^2 A T_6^5)^{{1\over6}}$$

Since most stellar reactions happen in a fairly narrow band of energies,
S(E) will have a nearly constant value over this band averaging to 
$S_0$. With this, the reaction rate per pair of particles, turns out to be:

$$<\sigma v> = \big[{8\over \pi\mu(kT)^3}\big]^{1/2} S_0 \int^{\infty}_0e^{-\tau -4({E-E_0\over \Delta})^2}dE = 4.5\times 10^{14} {S_0\over AZ_1Z_2} \tau^2 e^{-\tau} \rm cm^3s^{-1}$$
Here, 

$$\tau = 3 E_0 /kT = 42.5 (Z_1^2 Z_2^2 \mu /T_6)^{{1\over3}}$$ 
The maximum value
of the integrand in the above equation is: 

$$ \rm I_{max} = exp(-\tau)$$ 
The values
of $\rm E_0, I_{max}, \Delta,$ etc., apart from the Coulomb barrier for
several reactions are tabulated in Table \ref{tab: parameters}
for $T_6 = 15$. 

As the
nuclear charge increases, the Coulomb barrier increases, and the Gamow peak
$E_0$ also shifts towards higher energies. Note how rapidly the maximum
of the integrand $I_{max}$ decreases with the nuclear charge and the Coulomb
barriers. The effective width $\Delta$ is a geometric mean of $E_0$ and
kT, and $\Delta/2$ is much less rapidly varying between reactions (for
$kT \ll E_0$). The rapid variation of $I_{max}$  indicates that of several
nuclei present in the stellar core, those nuclear pairs will have the largest
reaction rates, which have the smallest Coulomb barrier.
The relevant nuclei will be consumed most rapidly at that stage. 
(Note however
that for the p+p reaction, apart from the Coulomb barrier, the strength of
the weak force, which transforms a proton to a neutron also
comes into play). 

When nuclei of the smallest Coulomb barrier are
consumed, there is a temporary dip in the nuclear generation rate, and the
star contracts gravitationally until the temperature rises to a point where
nuclei with the next lowest Coulomb barrier will start burning. At that
stage, further contraction is halted.
The star therefore goes through well defined stages
of different nuclear burning phases in its core at later epochs dictated
by the height of the Coulomb barriers of the fuels.
Note also from the Table \ref{tab: parameters}, 
how far $E_0$, the effective mean energy
of reaction is below the Coulomb barrier at the relevant temperature.
The stellar burning is so slow because the reactions are taking place at 
such a far sub-Coulomb region, and this is why the stars can last so long.

\begin{table}[hb] 
\vspace*{-12pt}
\caption[]{
Parameters of the thermally averaged reaction rates at $T_6 =15$.
}
\vspace*{-14pt}
\begin{center}
\begin{tabular}{llllll}
\hline\\[-10pt]
Reaction & Coulomb & Gamow & $I_{max}$  & {\phantom{$00$}}$\Delta$ & {\phantom{$0$}}$(\Delta)I_{max}$\\ 
         & Barrier & Peak ($E_0$) & ($e^{-3E_0/kT}$)    &           & \\
         & (MeV)   & (keV)   &                       & {\phantom{$0$}}(keV)      &\\
\hline\\[-10pt]
p + p     & 0.55& 5.9 & $1.1 \times 10^{-6}$ & 
{\phantom{$00$}}6.4 & $7 \times 10^{-6}$\\
p + N     & 2.27& 26.5 & $1.8 \times 10^{-27}$ & {\phantom{$0$}}13.6 & $2.5 \times 10^{-26}$\\ 
$\alpha$ + C$^{12}$ & 3.43& 56 & $3 \times 10^{-57}$ & {\phantom{$0$}}19.4 & $5.9 \times 10^{-56}$\\ 
O$^{16}$ + O$^{16}$ & 14.07& 237 & $6.2 \times 10^{-239}$ & 
{\phantom{$0$}}40.4 & $2.5 \times 10^{-237}$\\
\hline 
\end{tabular}
\end{center}
\label{tab: parameters}
\end{table}

The above discussion assumes that a bare nuclear Coulomb potential
is seen by the charged projectile. For nuclear reactions measured
in the laboratory, the target nuclei are in the form of atoms with
electron cloud surrounding the nucleus and giving rise to a screened
potential -- the total potential then goes to zero outside the atomic radius.
The effect of the screening is to reduce the effective height of the
Coulomb barrier. 
Atoms in the stellar interiors are in most cases in highly stripped state,
and nuclei are immersed in a sea of free electrons which tend to cluster
near the nucleus. When the stellar density increases, the so called
Debye-Huckel radius $R_D = (kT/ 4\pi e^2 \rho N_A \xi)^{1/2}$ , 
(here: $\xi = \sum_i (Z^2_i + Z_i) X_i/A_i$) which is
a measure of this cluster ``radius", decreases, and the effect of shielding
upon the reaction cross-section becomes more important.
This shielding effect enhances 
thermonuclear reactions inside the star. The enhancement factor
$f_0 = \rm exp (0.188 Z_1 Z_2 \xi \rho^{1/2} T_6^{-3/2}$,
varies between 1 and 2 for typical densities and compositions \cite{Sal54}
but can be large at high densities.

\section{Hydrogen burning: the pp chain}

The quantitative aspects of the problem of solar
energy production with details of known nuclear physics of converting  
hydrogen into helium
was first worked out by von Weizs\"acker (1937-38) \cite{Wei37}, \cite{Wei38}
and Bethe \& Critchfield (1938-1939) \cite{Bet38}, 
which made it clear that two different sets of 
reactions : the p-p chains and the CN cycle can do this conversion. 
This happens in the core of the star initially (at the ``main sequence"
stage), and then later in the 
life of a star in a shell of burning hydrogen around an inert core of He.

In the first generation of stars in the galaxy only the p-p cycle may have 
operated. In second generation, heavier elements like C, N from the ashes 
of burning in previous stars are available and they too can act as 
catalysts to have thermonuclear fusion of hydrogen to helium.
[A recent discovery (\cite{Chr02}) of a low-mass star with an 
iron abundance as
low as 1/200,000 of the solar value (compare the previous record
of lowest iron abundance less than 1/10,000 that of the sun), suggests
that such first generation stars are still around].

\begin{figure}[htb]
\vspace*{0.0cm}
%                 \insertplot{fig4.eps}
\plotone{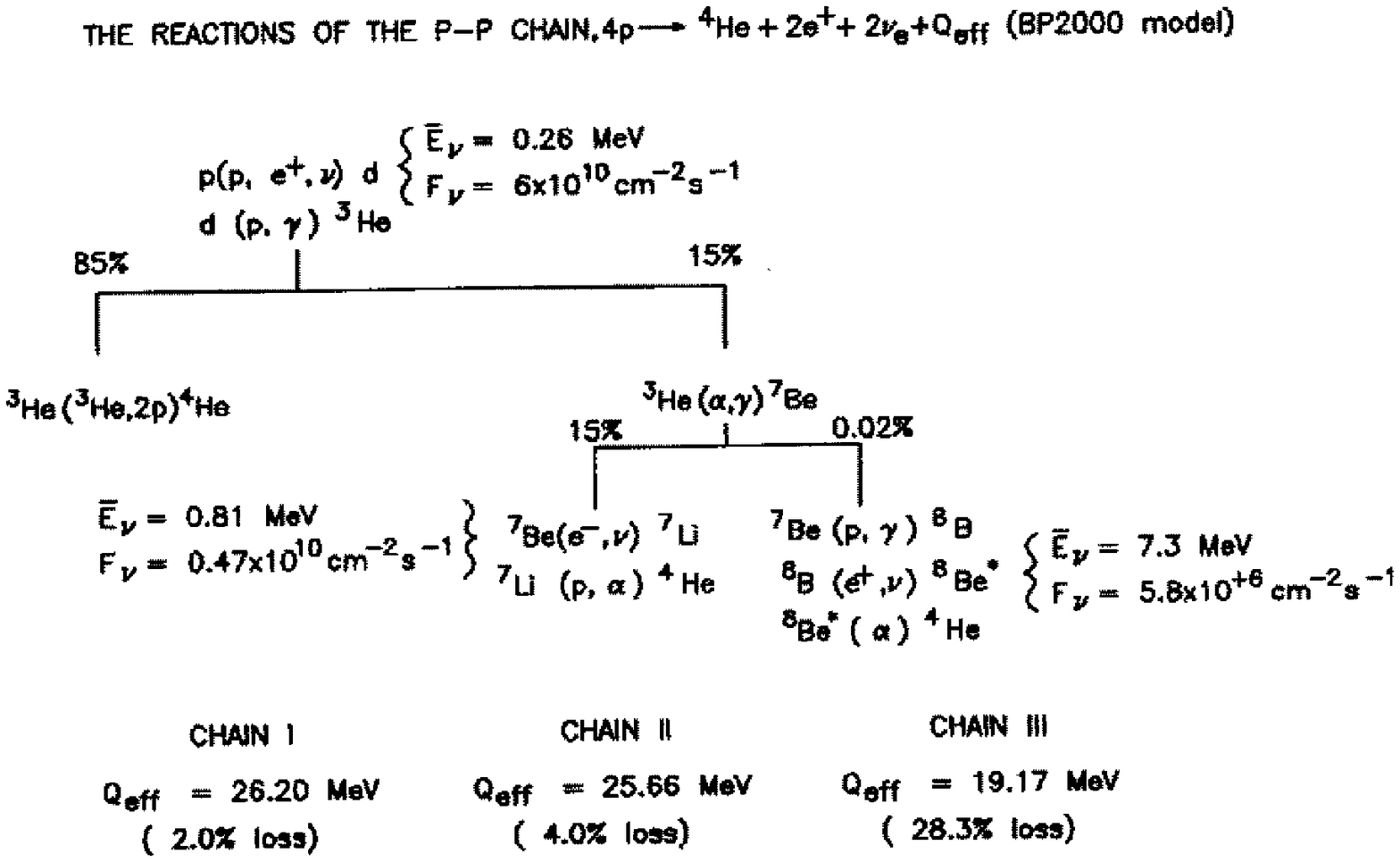}
\vspace*{-0.9cm}
\caption[]{
Reactions in the p-p chain start with the formation of deuterium
and $^3He$. Thereafter, the $^3He$ is consumed in the sun 85\% of
the time through the ppI chain, whereas the ppII and ppIII chains 
together account for 15\% of the time in the Bahcall Pinsonneault 2000
solar model. The ppIII chain occurs only $0.02 \%$ of the time,
but the $^8B$ $\beta^+$-decay provides the higher energy neutrinos
(average $\bar E_{\nu} = 7.3 \; \rm MeV$). The net result of the chains
is the conversion of four protons to a helium, with the effective
Q-values (reduced from 26.73 MeV) as shown, due to loss of energy
in escaping neutrinos.
}
\label{fig: ppchain}
\end{figure}

The sun with a central temperature of 15.6 million degrees, ($T^c_{6\odot}
= 15.6$) burns by p-p chains. Slightly more massive star (with 
central temperature $T_6  \geq 20$) burns H by also the CNO cycle. 
Davis et al.s' solar neutrino experiment \cite{Dav68}, which in 1968 had only
an upper limit of the neutrino flux, itself put a limit of
less than 9\% of the sun's energy is produced by the carbon-nitrogen cycle
(the more recent upper limit \cite{Bah03} is $7.3\%$, from an analysis
of several solar neutrino experiments, including the Kamland
measurements. Note however that
for the standard solar model, the actual contribution of CNO cycle
to solar luminosity is $\sim 1.5 \%$ \cite{Bah89}).
In CNO cycle, nuclei such as C, N, O serve as ``catalysts" do in a 
chemical reaction.
The pp-chain and the CNO cycle reaction sequences are illustrated
in Figs. \ref{fig: ppchain} and \ref{fig: CNOcycle}.

The pp-chain begins with the reaction $\rm p + p \rightarrow d + e^+ + \nu_e$.
Bethe and Critchfield \cite{Bet38} showed that weak nuclear reaction
is capable of converting a proton into a neutron during the brief
encounter of a scattering event. (This reaction overcomes the impasse
posed by the instability of $^2$He in the $\rm p + p \rightarrow ^2He$ reaction
via the strong or electromagnetic interactions, and as the next nucleus
$^3$Li reachable via these interactions is also unstable as a final product).
Since a hydrogen atom is less massive
than a neutron, such a conversion would be endothermic 
(requiring energy), except for the fact that a neutron in a deuterium nucleus
$^2$D can form a bound state with the proton with a binding energy of 
2.224 MeV, -- thus making the reaction exothermic with an available kinetic
energy of 0.42 MeV. The released positron soon pair annihilates and produces
photons which makes the total energy released to be 1.442 MeV.

Because of the low Coulomb barrier, in the p+p reaction ($E_c = 0.55$ MeV),
a star like the sun would have consumed all its hydrogen quickly
(note the relatively large value of $(\Delta) I_{max}$ in 
Table \ref{tab: parameters}),
were it not slowed down by the weakness of the weak interactions.
The calculation of probability of deuteron formation consists of two
separate considerations: 1) the penetration of a mutual potential
barrier in a collision of two protons in a thermal bath and 2) the
probability of the $\beta$-decay and positron and neutrino emission.
Bethe and Critchfield used the original Fermi theory (point interaction)
for the second part, which is adequate for the low energy process. 

\subsection{Cross-section for deuterium formation}  

The total Hamiltonian $H$ for the p-p interaction can be written as a
sum of nuclear term $H_n$ and a weak-interaction term $H_w$. As the weak
interaction term is small compared to the nuclear term, first order
perturbation theory can be applied and Fermi's ``Golden rule ", 
gives the differential cross-section as:

$$ d\sigma = {2\pi \rho(E) \over \hbar v_i} |<f | H_w | i>|^2 $$
here $\rho (E) = dN/dE$, is the density of final states in the interval $dE$
and $v_i$ is the relative velocity of the incoming particles.
For a given volume V, the number of states dn between p and p+dp is:-

$$ dN = dn_e dn_{\nu} = (V {4 \pi p_e^2 dp_e \over h^3}) (V {4 \pi p_{\nu}^2 dp_{\nu} \over h^3}) $$
By neglecting the recoil energy of deuterium (since this is much heavier
than the outgoing positron in the final state) and neglecting the mass of the
electron neutrino, we have: 
$E=E_e + E_{\nu}=E_e + c p_{\nu}$ and $dE=dE_{\nu}
= c p_{\nu}$, for a given $E_e$ and,

$$ \rho(E) = dN(E)/dE = dn_e (dn_{\nu}/dE) = 16 \pi^2 V^2 /(c^3 h^6) p_e^2 (E-E_e)^2 dp_e = \rho(E_e) dp_e $$
The matrix element that appears in the differential cross section, may be
written in terms of the initial state wave function $\Psi_i$ of the two protons
in the entrance channel and the final state wave function 
$\Psi_f$ as:

$$ H_{if} = \int [\Psi_d \Psi_e \Psi_{\nu}]^* H_{\beta} \Psi_i d\tau $$
If the energy of the electron is large compared to $Z\times \rm Rydberg$
(Rydberg $R_{\infty}= 2 \pi^2 m e^4 /ch^3$), then a plane wave approximation
is a good one: 
$\Psi_e = 1/(\sqrt V)\rm exp(i {\bf \vec{k_e}} . {\bf \vec{r}})$ 
where the wavefunction is normalised over volume V.
(For lower energies, typically 200 keV or less, the electron wave-function
could be strongly affected by nuclear charge (see \cite{Sch83}).
Apart from this, the final state wave function: $[\Psi_d \Psi_e \Psi_{\nu}]$
has a deuteron part $\Psi_d$ whose radial part rapidly vanishes
outside the nuclear domain ($R_0$), so that the integration need not
extend much beyond $r \simeq R_0$ (for example, the deuteron
radius $R_d = 1.7$ fm). Note that because of the Q-value
of 0.42 MeV for the reaction, the kinetic energy of the electron
($K_e \leq 0.42$ MeV) and the average energy of the neutrinos
($\bar E_{\nu} = 0.26$ MeV) are low enough so that for both
electrons and neutrino wavefunctions, the product 
$k R_0 \leq 2.2\times 10^{-3}$
and the exponential can be approximated by the first term of the
Taylor expansion:

$$\Psi_e= 1/(\sqrt V)\rm [1+i({\bf \vec{k_e}}.{\bf \vec{r}})] \sim 1/(\sqrt V)$$
and
$$\Psi_{\nu} \sim 1/(\sqrt V)$$
Then the expectation value of the Hamiltonian, given a strength of
interaction governed by coupling constant $g$ is:

$$H_{if} = \rm \int [\Psi_d \Psi_e \Psi_{\nu}]^* H_{\beta} \Psi_i d\tau = {g\over V} \rm \int [\Psi_d ]^* \Psi_i d\tau$$
The integration over $d\tau$ can be broken into a space part $M_{space}$ and
a spin part $M_{spin}$, so that the differential cross-section is:

$$d\sigma = {2\pi \over \hbar v_i} {16 \pi^2 \over c^3 h^6} g^2 M^2_{spin} M^2_{space} p_e^2 (E-E_e)^2 d p_e $$
Thus the total cross-section upto an electron energy of $E$ can be
obtained by integration as proportional to:
$$\int_0^E p_e^2 (E-E_e)^2 d p_e = {(m_e c^2)^5 \over c^3} \int_1^W (W_e^2 -1)^{1/2} (W-W_e)^2 W_e d W_e $$
where $W= (E+m_ec^2)/m_ec^2$.
The integral over W can be shown as:
$$f(W)=(W^2-1)^{1/2}[{W^4\over30}-{3W^2\over20}-{2\over15}]+{W\over4}\; \rm ln[W+(W^2-1)^{1/2}]$$
so that:
$$\sigma = { m_e^5 c^4 \over 2 \pi^3 \hbar^7} f(W) g^2 M_{space}^2 M_{spin}^2$$
At large energies, the factor $f(W)$ behaves as:
$$f(W) \propto W^5 \propto {1\over 30} E^5$$

The process that we are considering: $\rm p + p \rightarrow d + e^+ + \nu_e$,
the final state nucleus (deuterium in its ground state) has 
$J_f^{\pi} = 1^+$, with a predominant relative orbital angular momentum
$l_f = 0$ and $S_f = 1$ (triplet S-state). For a maximum probability
of the process, called the super-allowed transition, there are no changes
in the {\it orbital} angular momentum between the initial and final states
of the nuclei. Hence for super-allowed transitions, the initial two interacting
protons in the $\rm p+p$ reaction that we are considering
must have $l_i = 0$. Since the two protons are identical particles,
Pauli principle requires $S_i = 0$, so that the total wavefunction will be
antisymmetric in space and spin coordinates. Thus, we have a process:

$$ | S_i =0, l_i = 0 >  \rightarrow  | S_f =1, l_f = 0 > $$
This is a pure Gamow-Teller\footnote{
In the beta-decay {\it allowed} approximation, where we neglect the
variation of the lepton wavefunctions and the nuclear momentum
over the nuclear volume (this is equivalent to neglecting all total lepton
orbital angular momenta $L > 0$) the total angular momentum is
carried off by the lepton is just their total spin: i.e. $S=1$ or 0,
since each lepton has spin ${1\over2}$. When spins of the leptons in the
final state are antiparallel, $s_e + s_{\nu} = s_{tot} = 0$ the
process is the Fermi transition with Vector coupling constant
$g = C_V$ (example of a pure Fermi decay:
$^{14}O (J_i^{\pi} = 0^+) \rightarrow ^{14}N(J^{\pi}_f = 0^+)$).
When the final state lepton spins are parallel,
$s_e + s_{\nu} = s_{tot} =1$, the process is Gamow-Teller with $g=C_A$.
Thus, for the Fermi coupling, there is no change in the (total) angular
momentum between the initial and final states of the nuclei ($\Delta J=0$).
For the Gamow-Teller coupling, the selection rules are: $\Delta J =0$
or $\pm 1$ (but the possibility $\Delta J = 0$ cannot proceed
in this case between two states of zero angular momentum).
The size of the matrix element for a transition depends essentially
on the overlap of the wavefunctions in the initial and final states.
In the case of ``mirror pair" of nuclei (the nucleus $A_Z = (2Z+1)_Z$
is the mirror of the nucleus $(2Z+1)_{Z+1}$), the wavefunctions are
very much alike as can be shown through simple heuristic arguments
(\cite{Fer51}). For these nuclei, typical $ft$-values range from
$\sim 1000 - 5000$ and are called super-allowed
transitions. For super-allowed transitions, which have maximum
decay probabilities, there are no changes in the {\it orbital} angular
momentum between the initial and final states of the nuclei.
In the $\rm p \; + \; p \rightarrow D \; + e^+ + \; \nu_e$ reaction,
the initial proton state is antisymmetric to an interchange of
space and spin coordinates and the final deuteron is symmetric in
this respect (in fact when the two protons are in the S state (which
is most favourable for their coming together), their spins will be
antiparallel (a singlet state) whereas the ground state of the deuteron
is a triplet S state. If this were the complete description of the
exchange symmetry properties of the Gamow-Teller transition
(permitting a change of spin direction of the proton as it transforms
to a neutron, changing the total spin by one unit) advocated here
this would actually be forbidden. However in the use of configuration
space in beta-decay process one must include isotopic spin as well.
The $^1S$ state of the two protons is symmetric to exchange of this
coordinate, whereas the deuteron (consisting of both a proton and a
neutron) function is antisymmetric in this coordinate. In the complete
coordinate system the transition is from an initial antisymmetric state to
another antisymmetric final state accompanied by a positron emission
(\cite{BetCri38}).
} transition with coupling constant
$g= C_A$ (due to the axial vector component which can be obtained,
for example, in the pure GT transitions of 
$^6He(0^+) \rightarrow ^6Li(1^+)$ decay).

\begin{figure}[htb]
\vspace*{-0.5cm}
%                 \insertplot{fig5.eps}
\plotone{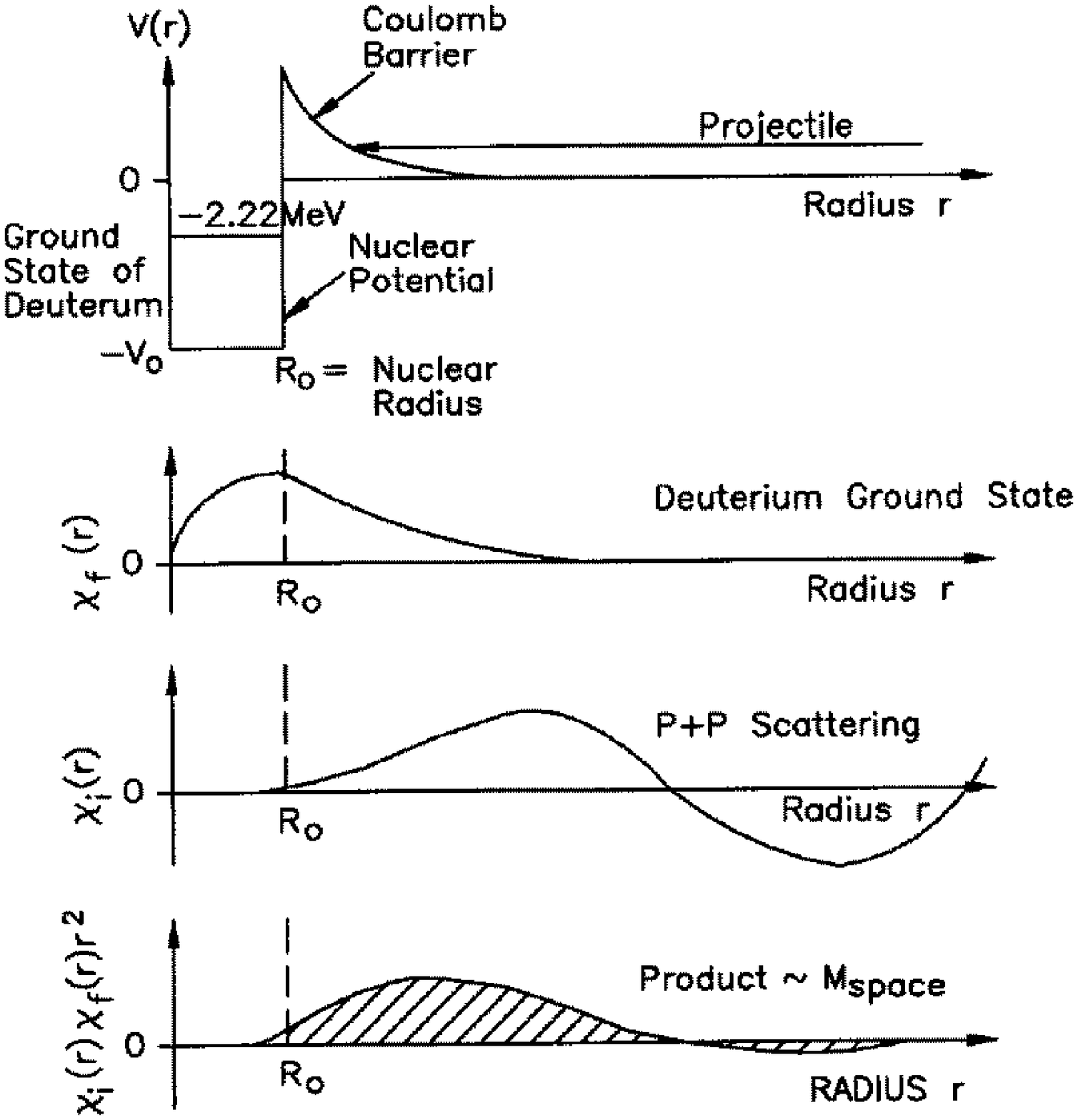}
\vspace*{-0.7cm}
\caption[]{
Schematic representation (after \cite{Rol88}) of the numerical calculation
of the spatial part of the matrix element $M_{space}$ in the 
$\rm p + p \rightarrow \; d + e^+ + \nu_e$ reaction. The top part
shows the potential well of depth $V_0$ and nuclear radius $R_0$ of
deuterium with binding energy of $-2.22 \rm \; MeV$. The next 
part shows the radius dependence of the deuterium radial wave
function $\chi_d(r)$. The wavefunction extends far outside the nuclear
radius with appreciable amplitude due to the loose binding of deuterium
ground state. The p-p wavefunction $\chi_{pp}(r)$
which comprise the $l_i = 0$ initial
state has small amplitude inside the final nuclear radius.
The radial part of the integrand entering into the calculation of
$M_{space}$ is a convolution of both $\chi_d$ and $\chi_{pp}$
in the second and third panels and is given with the hatched shading
in the bottom panel.
It has the major contribution far outside the nuclear radius. 
}
\label{fig: overlapintegral}
\end{figure}

The spin matrix element in the above expression for energy integrated
cross-section $\sigma$, is obtained from summing over the final states
and averaging over the initial states {\it and} dividing by 2 to take
into account that we have two identical particles in the initial state.
Thus,

$$\lambda = {1\over \tau} = {m^5 c^4 \over 2 \pi^3 \hbar^7 v_i} f(W) g^2 {M^2_{space} M^2_{spin} \over 2} $$
where, $M^2_{spin} ={(2J+1)\over(2J_1+1)(2J_2+1)} =3$. And the space matrix
element is:

$$M_{space} = \int_0^{\infty} \chi_f(r) \chi_i(r) r^2 dr$$
in units of $\rm cm^{3/2}$.
The above integral contains the radial parts of the nuclear
wavefunctions $\chi(r)$, and involves Coulomb wavefunctions for barrier
penetration at (low) stellar energies. The integral
has been evaluated by
numerical methods (\cite{Fri51}), and Fig. \ref{fig: overlapintegral} 
shows schematically
how the $M_{space}$ is evaluated for the overlap of
the deuterium ground state wavefunction with the initial
pair of protons state. (See also
\cite{Sal52}, \cite{Bah69}
for details of calculations
of the overlap integral and writing the astrophysical S-factor
in terms the beta decay rate of the neutron
\cite{Sal52} which takes into account of radiative corrections
to the axial-vector part of the neutron decay through an effective
matrix element, the assumption being that these are the same
as that for the proton beta decay in the pp reaction above).
In the overlap integral
one needs only the S-wave part
for the wavefunction of the deuteron $\psi_d$, 
as the D-wave part makes
no contribution to the matrix element \cite{Fri51}, although its contribution
to the normalisation has to be accounted for. The wavefunction
of the initial two-proton system $\psi_p$ is normalised to a plane
wave of unit amplitude, and again only the S-wave part is needed.
The asymptotic form of $\psi_p$ 
(well outside the range of nuclear forces) 
is given in terms of regular and
irregular Coulomb functions and has to be defined through quantities
related to the S-wave phase shifts in p-p scattering data).
The result is a miniscule total cross-section of $\sigma = 10^{-47} \rm cm^2$
at a laboratory beam energy of $E_p = 1 \; \rm MeV$, which cannot
be measured experimentally even with milliampere beam currents. 

The reaction $\rm p + p \rightarrow d + e^+ + \nu_e$ is a nonresonant
reaction and at all energies the rate varies smoothly with 
energy (and with stellar temperatures), with $S(0)=3.8\times 10^{-22}
\rm \; keV \; barn$ and $dS(0)/dE = 4.2\times10^{-24} \rm \; barn$.
At for example, the central temperature of the sun $T_6 =15$, this gives:
$<\sigma v>_{pp} = 1.2\times10^{-43} \rm\; cm^3 \; s^{-1}$
For density in the centre of the sun $\rho = 100 \; \rm gm\; cm^{-3}$
and and equal mixture of hydrogen and helium ($X_H = X_{He} =0.5$),
the mean life of a hydrogen nucleus against conversion to deuterium
is $\tau_H(H) = 1/N_H <\sigma v>_{pp} \sim 10^{10} \rm yr$. This is
comparable to the age of the old stars. The reaction is so slow
primarily because of weak interactions and to a lesser extent due
to the smallness of the Coulomb barrier penetration factor
(which contributes a factor $\sim 10^{-2}$ in the rate), and is
the primary reason why stars consume their nuclear fuel of hydrogen
so slowly.

\subsection{Deuterium burning}  

\begin{figure}[htb]
\vspace*{-0.5cm}
%                 \insertplot{fig6.eps}
\plotone{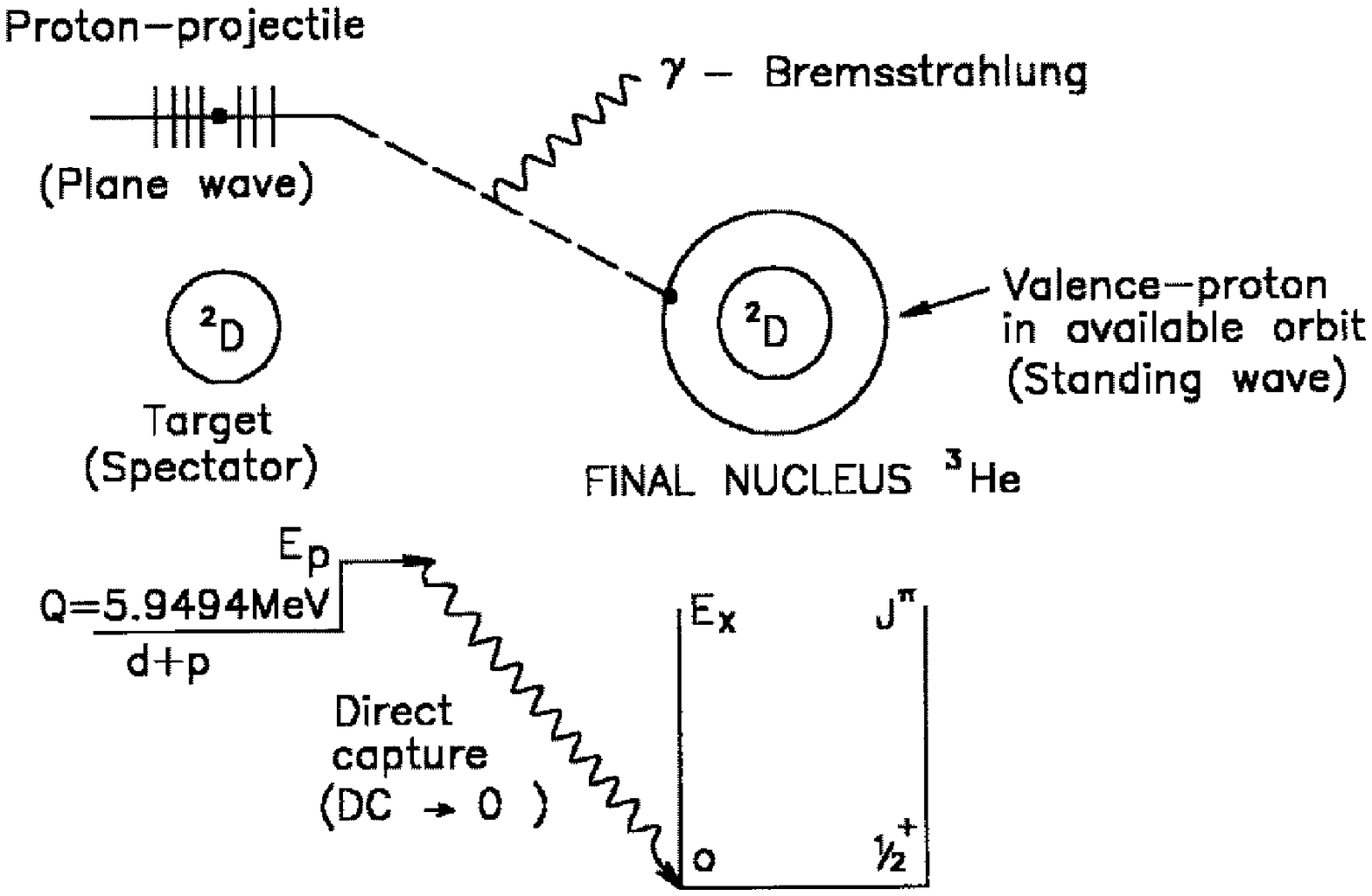}
\vspace*{-0.7cm}
\caption[]{
The Direct Capture reaction $d(p, \gamma)^3He$  to form $^3He$
in its ground state. The proton projectile (shown as a plane wave) 
radiates away a bremsstrahlung photon to be captured in a ``valence"
orbital around the $^2D$.
}
\label{fig: directcapture}
\end{figure}

Once deuterium is produced in the weak interaction mediated $\rm p + p$
reaction, the main way this is burnt in the
sun turns out to be:

$$d + p \rightarrow ^3He + \gamma $$
This is a nonresonant direct capture reaction to
the $^3He$ ground state with a Q-value of 5.497 MeV and
$S(0) = 2.5 \times 10^{-3} \rm keV \; barn$. 
The angle averaged cross-sections measured as a function
of proton + deuterium centre of mass energy, where the capture
transitions were observed in gamma-ray detectors at several
angles to the incident proton beam direction, are well explained
by the direct capture model (see Fig. \ref{fig: directcapture} after
\cite{Rol88}).

The reactions comprising the rest of
the (three) pp-chains start out with the predominant product of 
deuterium burning: $^3He$ (manufactured from $d + p$ reaction)
as the starting point.
The only other reactions with a $S(0)$ greater than the above
are: 
$\rm d(d,p)t \;, d(d,n)^3He \;, d(^3He,p)\\^4He$, and
$d(^3He, \gamma)^5Li$. However, because of the overwhelmingly
large number of protons in the stellar thermonuclear reactors,
the process involving protons on deuterium dominates. The rate
of this reaction is so fast compared to its precursor: 
$p+p \rightarrow d + e^+ \nu_e$, that the overall rate
of the pp-chain is not determined by this reaction.

One can show that the abundance ratio of deuterium to hydrogen
in a quasi-equilibrium has an extremely small value, signifying
that deuterium is destroyed in thermonuclear burning. The time 
dependence of deuterium abundance D is:

$$ {d D \over dt} = r_{pp} - r_{pd} = {H^2\over2} <\sigma v>_{pp} - HD <\sigma v>_{pd} $$
The self regulating system eventually reaches a state of quasi-equilibrium
and has:

$$(D/H) = <\sigma v>_{pp} / (2<\sigma v>_{pd}) = 5.6\times 10^{-18}$$ 
at $T_6 =5$ and $1.7\times 10^{-18}$ at $T_6 = 40$. For the solar system
however, this ratio is $1.5\times10^{-4}$ and the observed $(D/H)_{obs}$
ratio in the cosmos is $\sim 10^{-5}$. The higher cosmic ratio is due
to primordial nucleosynthesis in the early phase of the universe before
the stars formed. (The primordial deuterium abundance
is a key quantity used to determine the baryon density
in the universe). Stars only destroy the deuterium in their core
due to the above reaction.

\subsection{$\rm ^3He$ burning}  

The pp-chain-I is completed (see Fig. \ref{fig: ppchain})
through the burning of $^3He$ via the reaction: 

$$\rm ^3He \; + \; ^3He \rightarrow \;p \; + \; p \; + \; ^4He$$ 
with an S-factor: $S(0) = 5500 \; \rm \; keV \; barn$ and Q-value = 12.86 MeV.
In addition, the reaction:

$$\rm ^3He \; + \; D \rightarrow \; ^4He \; + p$$
has an S-factor: $S(0) = 6240 \; \rm keV \; barn$, but since the
deuterium concentration is very small as argued above,
the first reaction dominates the destruction of $^3He$
even though both reactions have comparable $S(0)$ factors.

$^3He$ can also be consumed by reactions with $^4He$ (the latter
is pre-existing from the gas cloud from which the star formed
and is synthesised in the early universe and in Pop III objects). 
These reactions proceed
through Direct Captures and lead to the ppII and ppIII parts
of the chain (happening $15\%$ of the time). Note
that the reaction $^3He(\alpha, \gamma)^7Be$ together with the
subsequent reaction: $^7Be(p, \gamma)^8B$ control the production
of high energy neutrinos in the sun and are particularly important
for the $^{37}Cl$ solar neutrino detector constructed by Ray Davis and
collaborators.

\subsection{Reactions involving $^7Be$}

As shown in Fig. \ref{fig: ppchain}, about 15\% of the time, $^3He$ is
burned with $^4He$ radiatively to $^7Be$. Subsequent reactions
involving $^7Be$ as a first step in alternate ways complete the fusion
process: $4 H \rightarrow  \; ^4He$ in the ppII and ppIII chains. 

\subsubsection{Electron capture process}

\begin{figure}[htb]
\vspace*{-0.5cm}
%                 \insertplot{fig7.eps}
\plotone{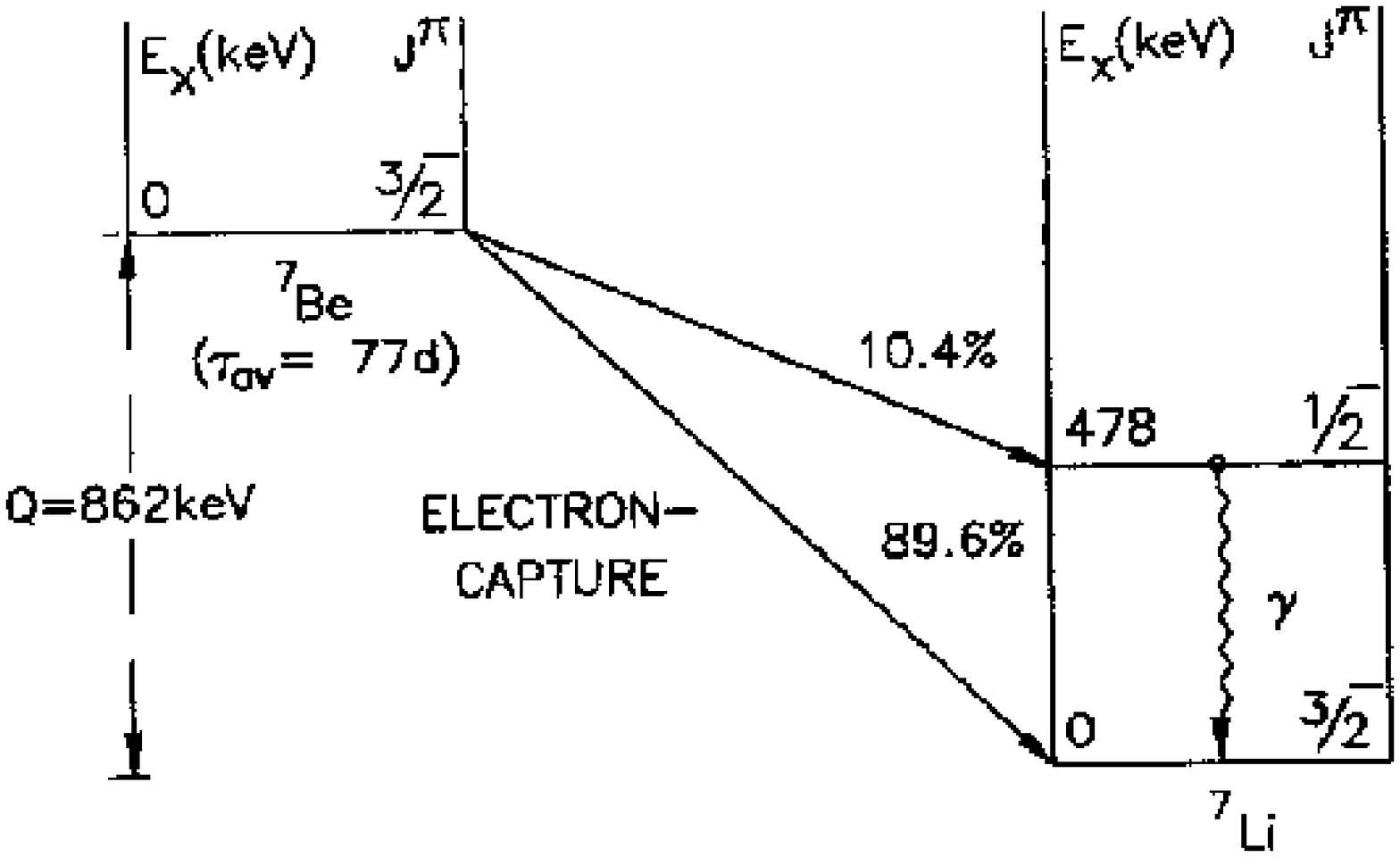}
\vspace*{-0.7cm}
\caption[]{
Electron capture on $^7Be$ nucleus. The capture proceeds 10.4\% of the time
to the first excited state of $^7Li$ at 478 keV, followed by a decay
to the ground state by the emission of a photon. The average energy of
the escaping neutrinos (which are from the ppII chain) is 814 keV.
}
\label{fig: ecapture7Be}
\end{figure}

The first step of the ppII chain is the electron capture reaction
on $^7Be$ : $^7Be \; + \; e^- \rightarrow \; ^7Li \; + \; \nu_e$ 
(see Fig \ref{fig: ecapture7Be}).
This decay goes both to the ground state of $^7Li$ as well as to its
first excited state at $E_X = 0.478 \; \rm keV, \; J^{\pi}={1\over2}^-)$ 
 -- the percentage of decays to the excited state being 10.4 \% in the 
laboratory. The energy 
released in the reaction with a Q-value of $0.862 \; \rm keV$
is carried away by escaping monoenergetic neutrinos with
energies: $E_{\nu} = 862$ and 384 keV. The measured laboratory mean life 
of the decay is $\tau = 76.9 \rm d$. 
The capture rate in the laboratory can be
obtained from Fermi's Golden Rule and utilising the fact
that the wavefunctions of both the initial nucleus and the final one
vanish rapidly outside the nuclear domain and the electron wavefunction
in that domain can be approximated as its value at $r=0$
and the neutrino wavefunction by a plane wave normalised to volume
V, so that $H_{if} = \Psi_e(0) g / \sqrt V \int \Psi^*_{^7Li} \Psi^{}_{^7Be} d\tau
= \Psi_e(0) g M_n/\sqrt V $, where $M_n$ represents the nuclear matrix
element and the resultant capture rate is:

$$ \lambda_{EC} = 1/\tau_{EC} = (g^2 M_n^2 /\pi c^3 \hbar^4) E_{\nu}^2 |\Psi_e(0)|^2 $$
In the laboratory capture process, any of the various electron shells
contribute to the capture rate; however the K-shell gives the dominant
contribution. At temperatures inside the sun, e.g. $T_6 = 15$,
nuclei such as $^7Be$ are largely ionised. The nuclei however
are immersed in a sea of free electrons resulting from the ionised
process and therefore electron capture from continuum states is possible 
(see e.g., \cite{Bet36}, \cite{Bah69b}). 
Since all factors in the capture of continuum
electrons in the sun are approximately the same as those in the
case of atomic electron capture, except for the respective electron
densities, the $^7Be$ lifetime in a star, $\tau_s$ is related
to the terrestrial lifetime $\tau_t$ by:

$${\tau_{fr} \over \tau_t} \sim {2 |\Psi_t(0)|^2 \over |\Psi_{fr}(0)|^2} $$
where $|\Psi_{fr}(0)|^2$ is the density of the free electrons 
$n_e = \rho/ m_H$ at
the nucleus, $\rho$ being the stellar density. 
The factor of 2 in the denominator takes care of the two
spin states of calculation of the $\lambda_t$ whereas the corresponding
$\lambda_{fr}$ is calculated by averaging over these two orientations.
Taking account of distortions of the electron wavefunctions due to
the thermally averaged  Coulomb interaction with nuclei of charge Z
and contribution due to hydrogen (of mass fraction $X_H$) and heavier nuclei,
one gets the continuum capture rate as: 

$$\tau_{fr} = { 2|\Psi_t(0)|^2 \tau_t \over (\rho/M_H)[(1+X_H)/2] 2\pi Z \alpha (m_e c^2 /3 kT)^{1/2}}$$
with $|\Psi_e(0)|^2 \sim (Z/a_0)^3/ \pi$. Bahcall et al \cite{Bah69}
obtained for the $^7Be$ nucleus a lifetime:

$$\tau_{fr} (^7Be) = 4.72 \times 10^8 {T^{1/2}_6 \over \rho(1+X_H)} \rm s$$ 
The temperature dependence comes from the nuclear Coulomb field corrections
to the electron wavefunction which are thermally averaged. For solar condition
the above rate \cite{Bah69b} gives a continuum capture rate of
$\tau_{fr}(^7Be) = 140 d$ as compared to the terrestrial mean life of $\tau_t =
76.9 d$.  Actually, under stellar conditions, there is a partial
contribution from some $^7Be$ atoms which are only partially ionised,
leaving electrons in the inner K-shell. So the contributions of
such partially ionised atoms have to be taken into account.
Under solar conditions the K-shell electrons from partially ionised
atoms give another 21\% increase in the total decay rate. Including this,
gives the solar lifetime of a $^7Be$ nucleus as: $\tau_{\odot} (^7Be) = 120 d$.
In addition, the solar fusion reactions have to be corrected for
plasma electrostatic screening enhancement
effects. For a recent discussion of the issues see \cite{Bah02}.

\subsubsection{Capture reaction leading to $^8B$}

Apart from the electron capture reaction, the $^7Be$ that is produced 
is partly consumed by proton capture via: $^7Be (p, \alpha) ^8B$ reaction.
Under solar conditions, this reaction happens
only $0.02 \%$ of the time. The proton capture on $^7Be$ proceeds at
energies away from the 640 keV resonance via the direct capture process.
Since the product $^7Li$ nucleus emits an intense $\gamma$-ray flux of 478 keV,
this prevents the direct measurement of the direct capture to
ground state $\gamma$-ray yield. The process is studied indirectly
by either the delayed positron or the breakup of the
product $^8B$ nucleus into two alpha particles. 
This reaction has a weighted average $S(0) = 0.0238 \rm \; keV barn$
\cite{Fil83}.

\begin{figure}[htb]
\vspace*{-0.5cm}
%                 \insertplot{fig8.eps}
\plotone{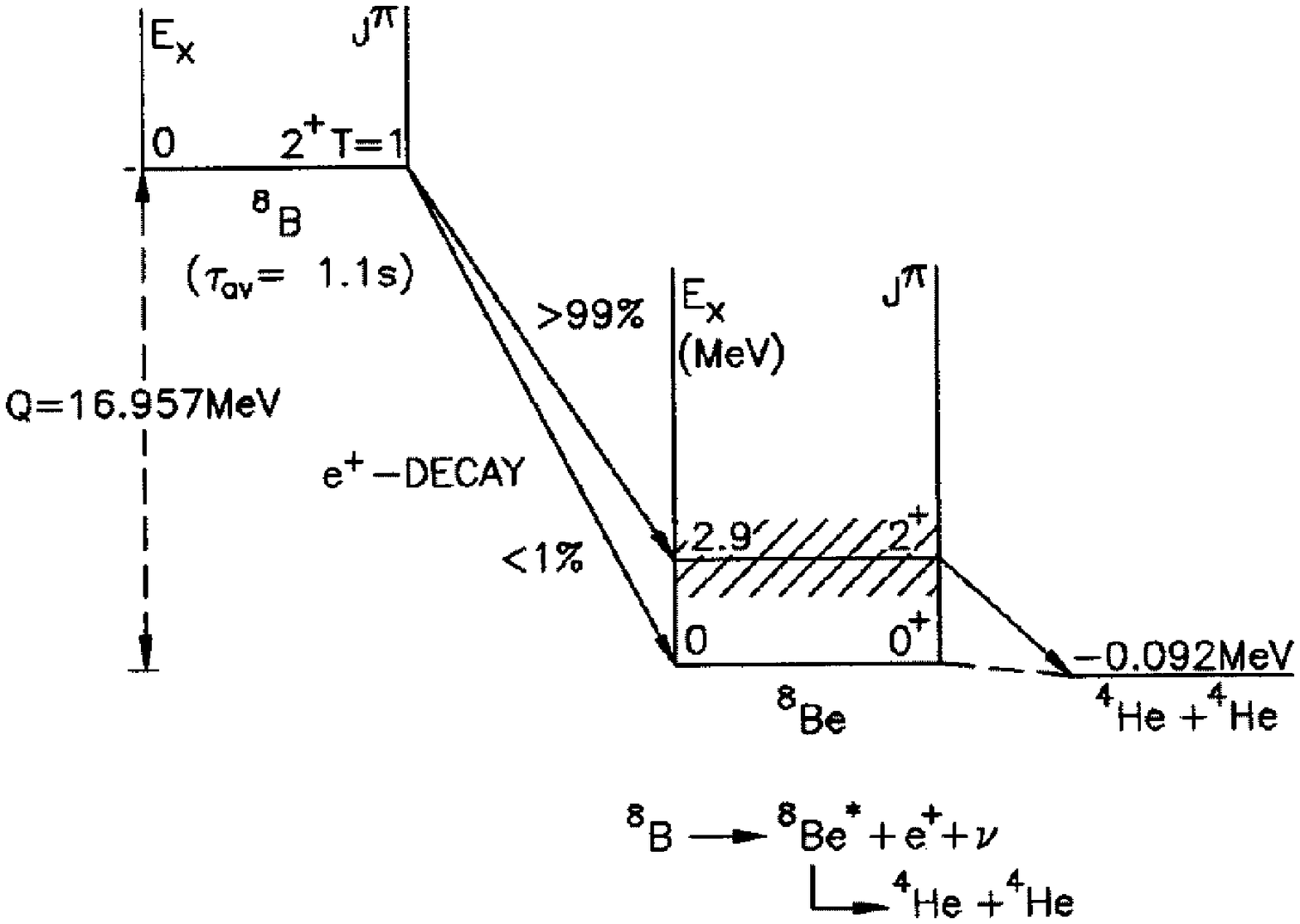}
\vspace*{-0.7cm}
\caption[]{
The decay scheme of $^8B$ with positron emission, which goes
to the first excited state of $^8$Be at $E_X = 2.9 \; \rm MeV$
with a width of $\Gamma = 1.6 \; \rm MeV$. The $^8Be$ nucleus 
itself fissions into two alpha particles. The neutrinos accompanying
the positron decay of $^8B$ are the higher energy solar neutrinos
with $\bar E_{\nu} = 7.3 \; \rm MeV$.
}
\label{fig: B8decay}
\end{figure}

The product: $^8B$ is a radioactive nucleus that decays with a lifetime
$\tau = 1.1$ s: % and $Q=17.979$ MeV:

$$^8B \rightarrow ^8Be + e^+ + \nu_e $$
The positron decay of $^8B (J^{\pi} = 2^+)$ 
goes mainly to the $\Gamma = 1.6$ MeV
broad excited state in $^8Be$ at excitation energy $E_x= 2.94$ MeV 
($J^{\pi} = 2^+$) due to the selection rules (see Fig. \ref{fig: B8decay}). 
This excited state
has very short lifetime and quickly decays into two $\alpha$-particles.
This completes the ppIII part of the pp-chain. The average energy of
the neutrinos from $^8B$ reactions is: $\bar E_{\nu}(^8B) = 7.3 $ MeV.
These neutrinos, having relatively high energy, play an important
role in several solar neutrino experiments.

\section{The CNO cycle and hot CNO}  

\begin{figure}[htb]
\vspace*{-0.5cm}
%                 \insertplot{fig9.eps}
\plotone{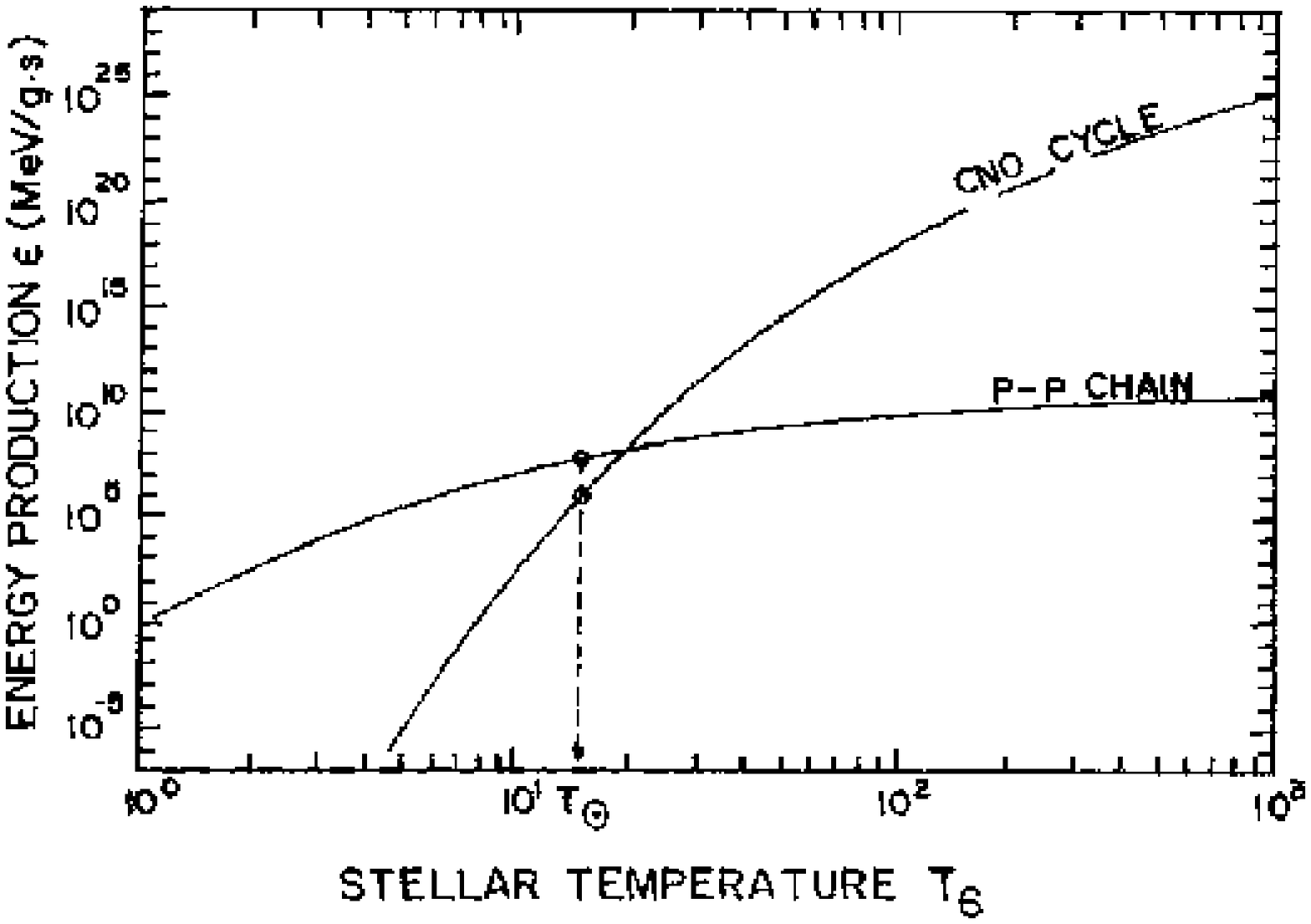}
\vspace*{-0.7cm}
\caption[]{
Comparison of the temperature dependence of the p-p chain
and the CNO cycle energy production. The points marked
for the solar central temperature $T_{\odot} = T_6 =15$
are shown on both graphs. The CNO cycle generation dominates
over the pp-chain at temperatures higher than $T_6 =20$,
so that for sun like stars, the pp-chain dominates. For more
massive stars, the CNO cycle dominates as long as one of
the catalysts: C, N, or O have initial mass concentration
at least 1\%. Note the logarithmic scales of the graph and how
both rates drop sharply with decreasing temperature, with that of
CNO cycle even more drastic due to higher Coulomb barriers.
}
\label{fig: CNOvsPP}
\end{figure}

The sun gets most of its energy generation through the pp-chain reactions
(see Fig. \ref{fig: CNOvsPP}). However, as the central temperature
(in stars more massive than the sun) gets higher, the CNO cycle (see
below for reaction sequence) comes to dominate over the pp-chain at
$T_6$ near 20 (this changeover assumes the solar CNO abundance,
the transition temperature depends upon CNO abundance in the star). 
The early generation of stars (usually referred to as 
the Population II (or Pop II stars), although there is an even
earlier generation of Pop III stars)
generated energy primarily through the pp-chain. These stars are still
shining in globular clusters, and being of mass lower than that of the
sun, are very old. Most other stars that we see today
are later generation stars formed from the debris of heavier
stars that contained heavy elements apart from (the most abundant) hydrogen.
Thus in the second and third generation stars (which are slightly
heavier than the sun) where higher central temperatures are possible because
of higher gravity, hydrogen burning can take place through faster chain of
reactions involving heavy elements C, N, and O which have some reasonable
abundance (exceeding 1\%)) compared to other heavy elements like Li, Be, B
which are extremely low in abundance. The favoured reactions involve
heavier elements (than those of the pp-chain) which have the
smallest Coulomb barriers
but with reasonably high abundance. Even though the Coulomb barriers
of Li, Be, B are smaller than those of C, N, O (when protons are the
lighter reactants (projectiles)), they lose out due to their lower abundance.

In 1937-1938, Bethe and von Weizs\"acker independently suggested the
CN part of the cycle, which goes as: 

$$^{12}C(p,\gamma)^{13}N(e^+\nu_e)^{13}C(p,\gamma)^{14}N(p,\gamma)^{15}O(e^+\nu)^{15}N(p,\alpha)^{12}C$$ 
This has the net result, as before: $4 p \rightarrow 
^4He + 2 e^+ + 2\nu_e$ with a $Q=26.73$. In these reactions, the
$^{12}C$ and $^{14}N$ act merely as catalysts as their nuclei are
``returned" at the end of the cycle. Therefore the $^{12}C$ nuclei
act as seeds that can be used over and over again, even though
the abundance of the seed material is miniscule compared
to the hydrogen. But note that there is a loss
of the catalytic material from the CN cycle that takes place through the 
$^{15}N (p, \gamma)^{16}O$ reactions. However, the catalytic material is
subsequently returned to the CN cycle by the reaction:
$^{16}O(p, \gamma) ^{17}F(e^+\nu_e)^{17}O(p,\alpha)^{14}N$.

\begin{figure}[htb]
\vspace*{-0.5cm}
%                 \insertplot{fig10.eps}
\plotone{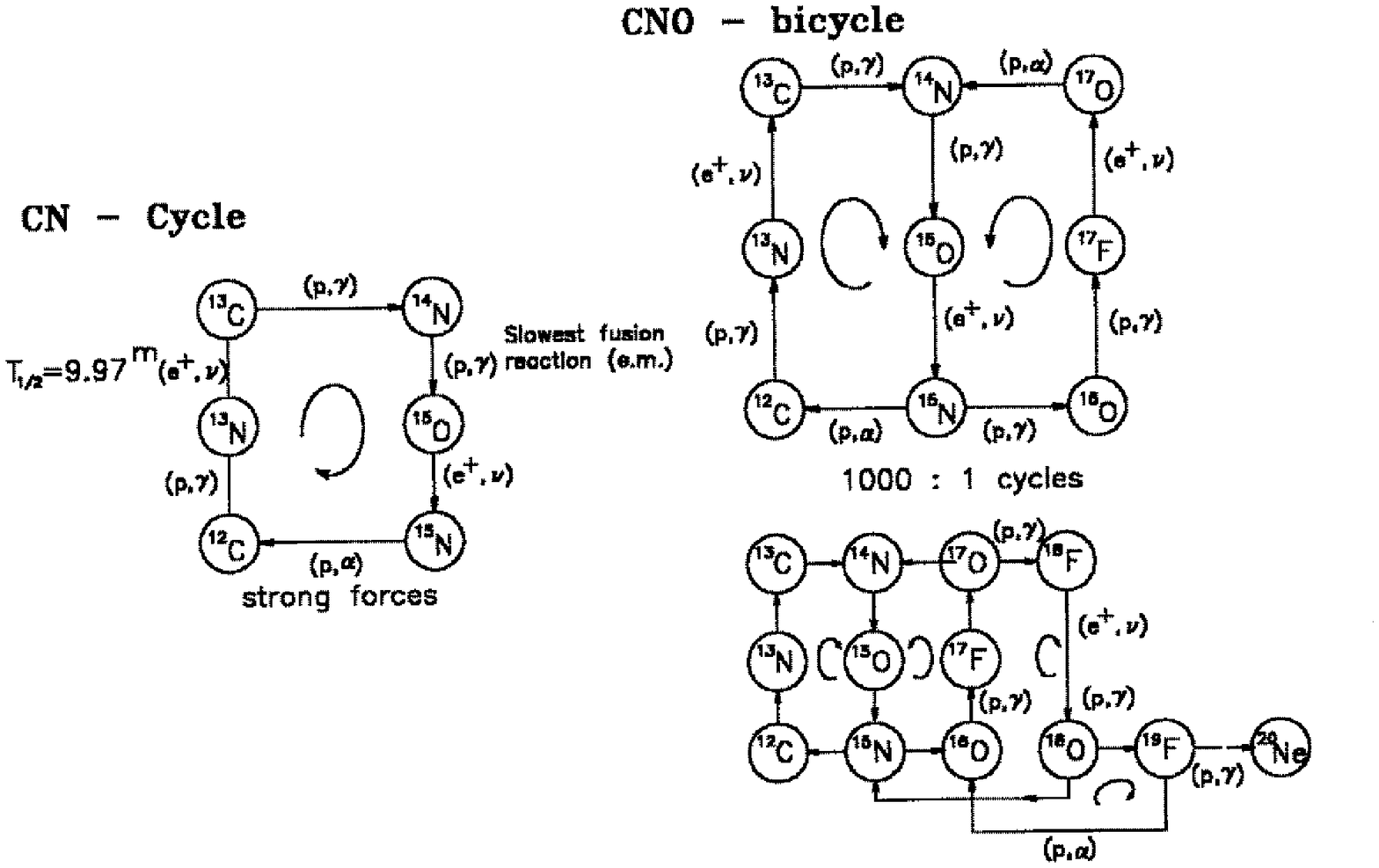}
\vspace*{-0.7cm}
\caption[]{
The various CNO cycles. The left part is the CN cycle where only
C and N serve as catalysts for the conversion of four protons into
$^4He$. Here the slowest fusion reaction is (p,$\gamma$) reaction
on $^{14}N$ whereas the slower $\beta$-decay has a half-life of 
$9.97 \rm m$. In the CNO bi-cycle (right part), 
there is leakage from the CN cycle
to the ON cycle through the branching at $^{15}N$. The flow is returned
to the CN cycle (which cycles 1000 times for each ON cycle) through
$^{17}O(p, \alpha)^{14}N$. The right bottom part represents additional
cycles linking into the CNO cycle through the $^{17}O(p, \gamma)^{18}F$ 
reaction \cite{Rol88}.
}
\label{fig: CNOcycle}
\end{figure}

In the CN cycle (see Fig \ref{fig: CNOcycle}), the two neutrinos involved
in the beta decays (of $^{13}N$ ($t_{1/2} = 9.97 \rm min$)
and $^{15}O$ ($t_{1/2} = 122.24 \rm s$))
are of relatively low energy and most of the total energy $Q=26.73$ MeV
from the conversion of four protons into helium is deposited in the
stellar thermonuclear reactor. The rate of the energy production is
governed by the slowest thermonuclear
reaction in the cycle. Here nitrogen isotopes
have the highest Coulomb barriers in charged particle reactions, 
because of their $Z=7$. Among them $^{14}N(p, \gamma)^{15}O$ is the 
slowest because this reaction having a final state photon is
governed by electromagnetic forces while that involving the
other nitrogen isotope: $^{15}N(p, \alpha)^{12}C$ is governed by 
strong forces and is therefore faster.

From the CN cycle, there is actually a branching off from $^{15}N$
by the reaction $^{15}N (p, \gamma)^{16}O$ mentioned above. This 
involves isotopes of oxygen, and is called the ON cycle;
finally the nitrogen is returned to the CN cycle through $^{14}N$.
Together the CN and the ON cycles, constitutes the CNO bi-cycle.
The two cycles differ considerably in their relative cycle-rates:
the ON cycle operates only once for every 1000 cycles of the
main CN cycle. This can be gauged from the S(0) factors of the
two sets of reactions branching off from $^{15}N$: for the
$^{15}N(p, \alpha)^{12}C$ reaction $S(0)= 65 \; \rm MeV \; b$,
whereas for $^{15}N(p, \gamma)^{16}O$, it is $64 \rm \; keV \; b$,
i.e. a factor of 1000 smaller.

\subsection{Hot CNO and rp-process}

\begin{figure}[htb]
\vspace*{-0.5cm}
%                 \insertplot{fig12.eps}
\plotone{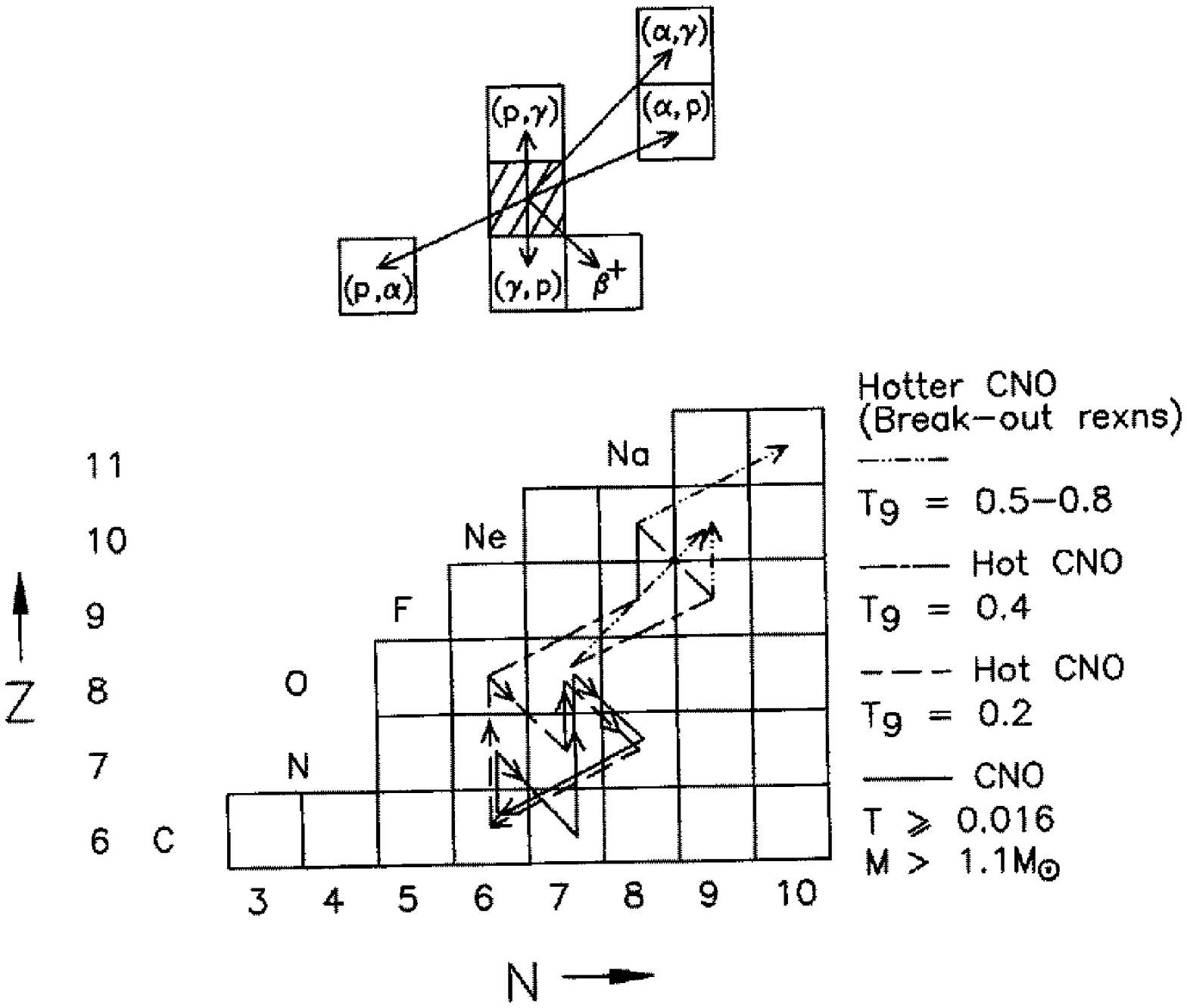}
\vspace*{-0.7cm}
\caption[]{
A given nucleus in the CNO cycles can participate in a variety
of particle capture or emission reactions. The path in the
$(N,Z)$ plane for the CNO, hot CNO and breakout to rp-process
are shown for successively higher temperatures.
}
\label{fig: NZplane-hotCNO}
\end{figure}

The above discussion of CNO cycle is relevant for typical temperatures
$T_6 \geq 20$. These are found in quiescently hydrogen burning stars with 
solar composition which are only slightly more massive than the sun. 
There are situations where the hydrogen burning takes place at
temperatures ($T \sim 10^8 - 10^9 \rm \; K$) which are far in excess
of those found in the interiors of the ordinary ``main sequence" stars.
Examples of these are: hydrogen burning at the accreting surface of a
neutron star or in the explosive burning on the surface
of a white dwarf, i.e. novae, or the outer layers of
a supernova shock heated material in the stellar mantle. 
These hot CNO cycles operate under such
conditions on a rapid enough timescale (few seconds) so that even
``normally" $\beta$-unstable nuclei like $^{13}N$ will live long enough
to be burned by thermonuclear charged particle reactions, before
they are able to $\beta$-decay. The process of normal CNO and a typical
part of hot CNO (at $T_9 = 0.2$) in the (N,Z) plane is illustrated
in Fig. \ref{fig: NZplane-hotCNO}. So, unlike the normal CNO
the amount of hydrogen to helium conversion in hot CNO is limited by the
$\beta$-decay lifetimes of the proton-rich nuclei like: $^{14}O$ and
$^{15}O$ rather than the proton capture rate of $^{14}N$. Wallace and Woosley
(1981, \cite{Wal81}) has shown that for temperatures, $T \geq 5 \times 10^8 K$, 
nucleosynthesised material can leak out of the cycles. This leads to
a diversion from lighter to heavier nuclei and is known as the rapid
proton capture or rp-process. 

\begin{figure}[htb]
\vspace*{-0.5cm}
%                 \insertplot{fig11.eps}
\plotone{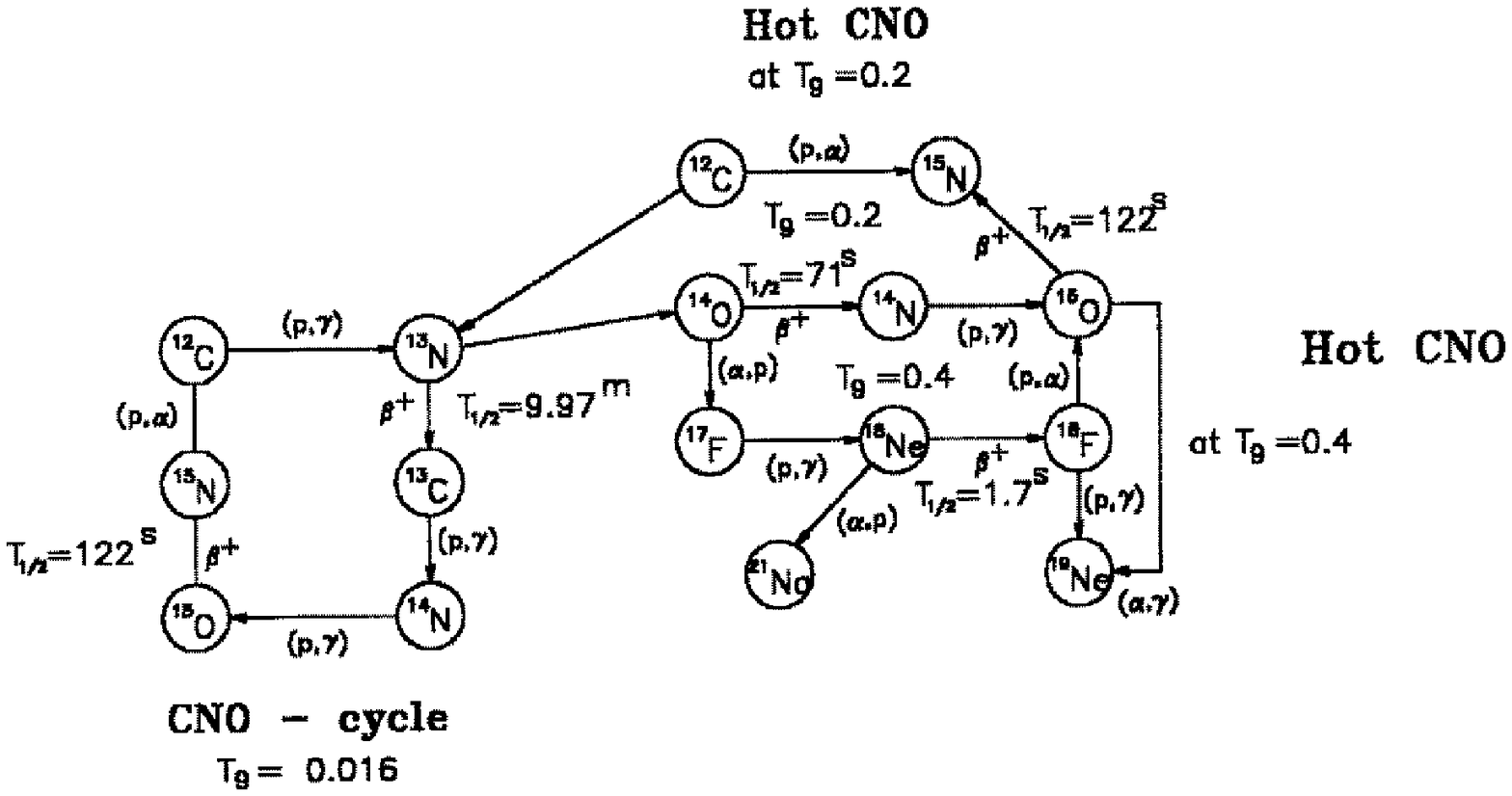}
\vspace*{-0.7cm}
\caption[]{
The ``hot CNO" reaction schemes. While the ``normal" CNO operates around
$T_9 = 0.016$, at higher temperatures $T_9 \sim 0.2$, proton capture
on $^{13}N$ can begin to compete with $\beta$-decay and the hot CNO
ensues. At even higher temperatures $T_9 \sim 0.4$, reactions that break 
out of the CNO cycle compete. These breakouts are the beginnings of the
rp-process.
}
\label{fig: hotCNO}
\end{figure}

\begin{figure}[htb]
\vspace*{-0.5cm}
%                 \insertplot{fig13.eps}
\plotone{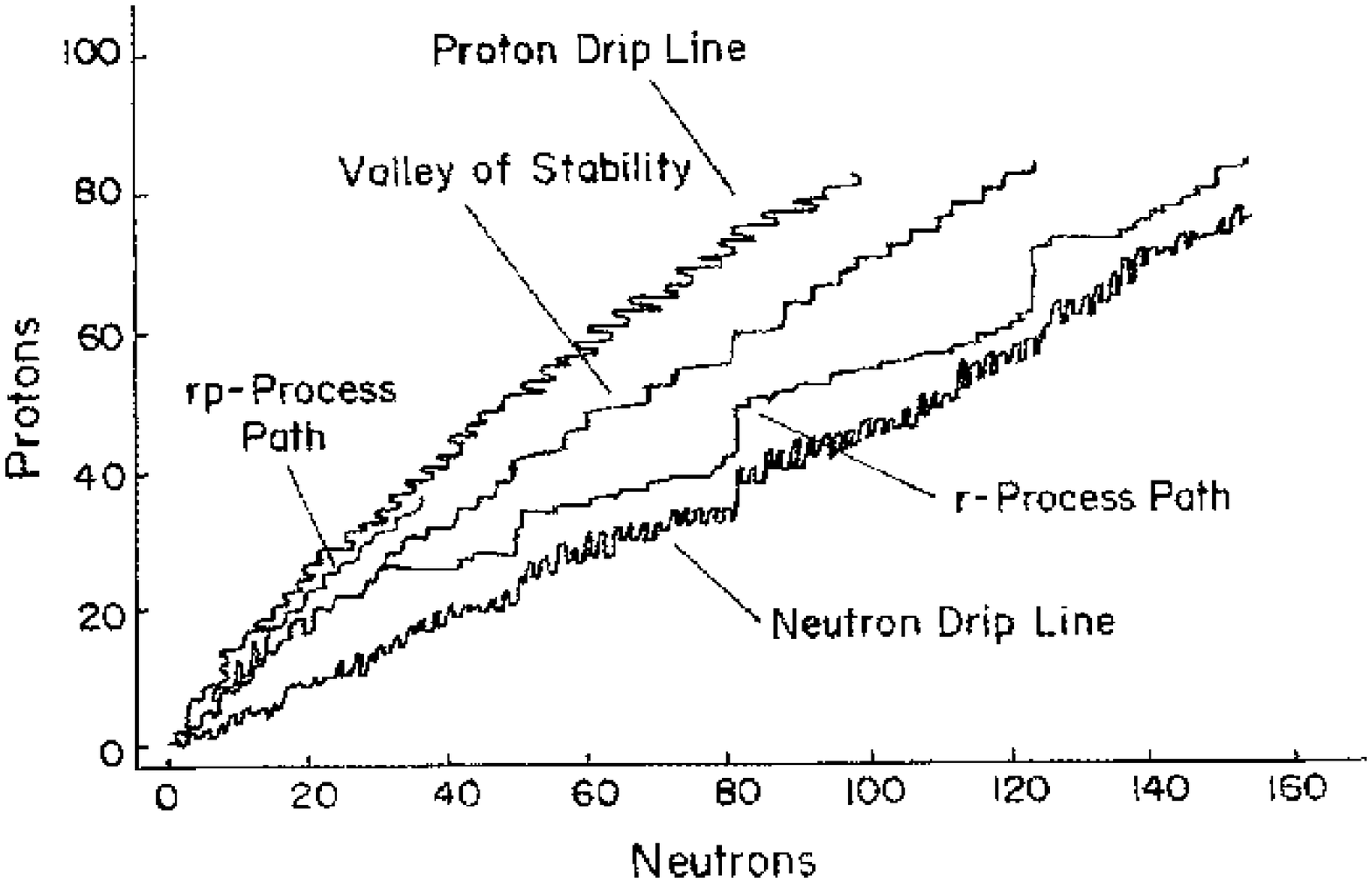}
\vspace*{-0.7cm}
\caption[]{
Schematic paths of r-process and rp-process in the (N,Z) plane
with respect to the valley of beta-stability, the neutron
drip and proton-drip lines.
}
\label{fig: rp-r-proc}
\end{figure}

The nucleosynthesis path of rp-process of rapid proton addition
is analogous to the r-process of neutron addition.
The hot hydrogen bath converts CNO nuclei into isotopes near the region
of proton unbound nuclei (the proton drip line). For each neutron number, 
a maximum mass number A is reached where the proton capture must wait
until $\beta^+$-decay takes place before the buildup of heavier nuclei
(for an increased neutron number) can take place. Unlike the r-process
the rate of the rp-process is increasingly hindered due to the increasing
Coulomb barrier of heavier and higher-Z nuclei to proton projectiles.
Thus the rp-process does not extend all the way to the proton drip line
but runs close to the beta-stability valley and runs through
where the $\beta^+$-decay rate compares favourably with the proton captures.
A comparison of the reaction paths of rp- and r-processes in the
(N,Z) plane is given in Fig. \ref{fig: rp-r-proc}.
A useful web reference for the rp-process in a nutshell is Guidry (1994)
\cite{Gui94} (see also \cite{vWormer94}, \cite{Cha92}).

\section{Helium burning and the triple-$\alpha$ reaction}

After hydrogen burning in the core of the star has exhausted its fuel,
the helium core contracts slowly. Its density and temperature goes up
as gravitational energy released is converted to internal kinetic
energy. The contraction also heats hydrogen at the edge of the helium core,
igniting the hydrogen to burn in a shell. At a still later stage in
the star's evolution, the core has contracted enough to reach central
temperature density conditions: $T_6 = 100 -200$ and
$\rho_c = 10^2 - 10^5 \; \rm gm \; cm^{-3}$ when the stellar core settles down
to burn $^4He$ in a stable manner.  The product of helium burning
is $^{12}C$. Since in nature, the
$A=5$ and $A=8$ nuclei are not stable, the question arises as to how
helium burning bridges this gap. A direct interaction of three $\alpha$
particles to produce a $^{12}C$ nucleus would seem at first sight, to be too
improbable (as was mentioned, for example, in Bethe's 1939 paper \cite{Bet39}, 
which was primarily on the
CN cycle). However, \"Opik \cite{Opi51} and Salpeter \cite{Sal52}, \cite{Sal57}
independently proposed
a two step process where in the first step,
two $\alpha$ particles interact to produce
a $^8Be$ nucleus in its ground state (which is unstable to $\alpha$-breakup),
followed
by the unstable nucleus interacting with another $\alpha$-particle
process to produce a $^{12}C$ nucleus.

\begin{figure}[htb]
\vspace*{-0.5cm}
%                 \insertplot{fig14.eps}
\plotone{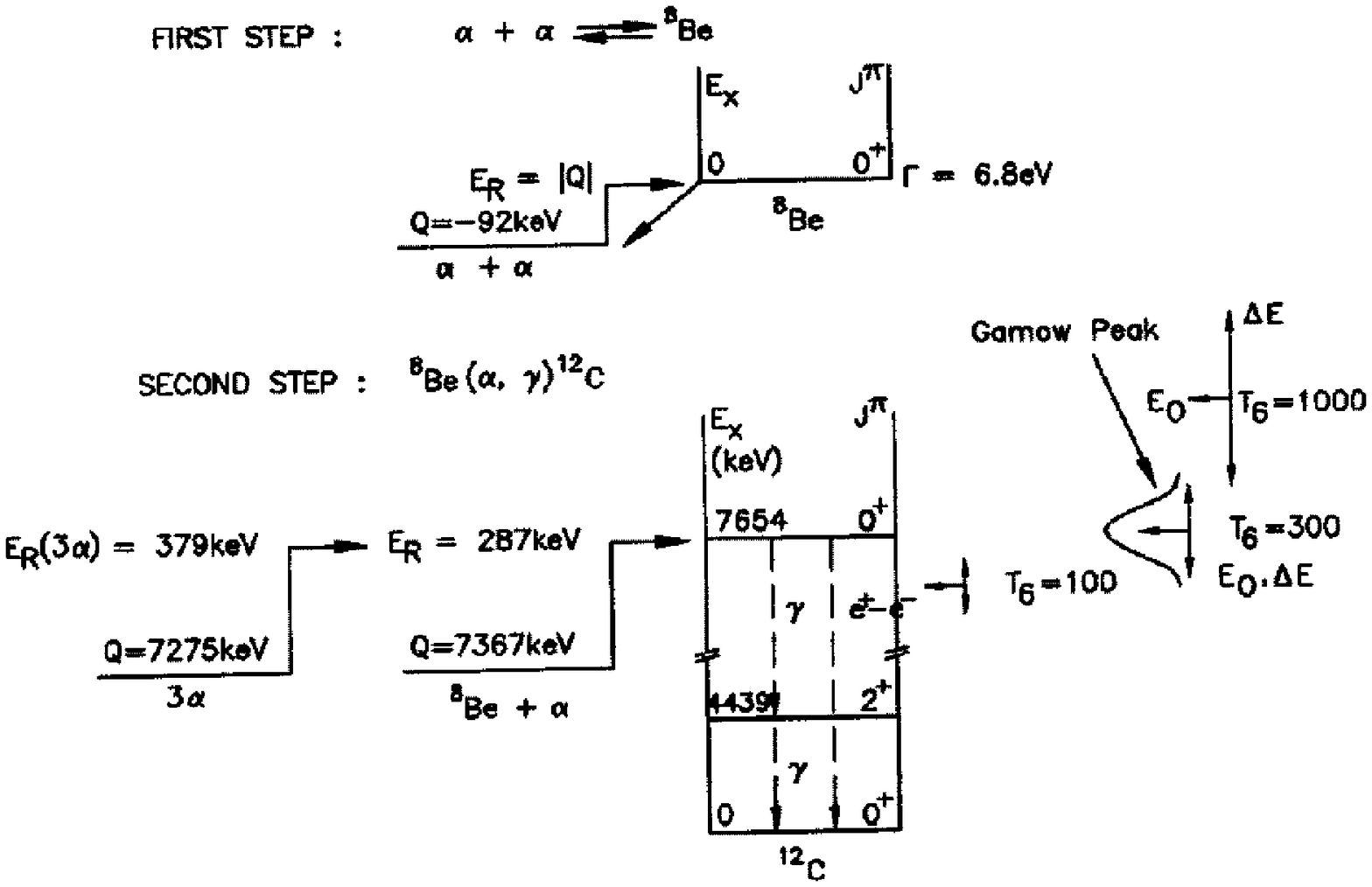}
\vspace*{-0.7cm}
\caption[]{
The triple alpha process of synthesising $^{12}C$ nucleus.
In the first step a small amount of $^8Be$ nuclei build-up 
in equilibrium with its decay products (both forward and backward
reactions involve alpha particles). The second step involves a
capture of another alpha particle by the unstable $^8Be$ nucleus
which proceeds via an s-wave resonance state in the product
nucleus $^{12}C$ which is located close to the Gamow energy
regions for temperatures indicated schematically 
by the three-way arrows on
the right.
}
\label{fig: 3alpha}
\end{figure}

Thus the triple alpha reaction begins with the formation of $^8Be$
that has a lifetime of only $1\times 10^{-16}$ s (this is found
from the width $\Gamma = 6.8$ eV of the ground state and is the cause of
the $A=8$ mass gap). This is however long compared to the transit time
$1\times 10^{-19}$ s of two $\alpha$-particles to scatter past each other 
non-resonantly with
kinetic energies comparable to the Q-value of the reaction namely,
$Q= -92.1 \rm \; keV$. So it is possible to have an equilibrium build-up
of a small quantity of $^8Be$ in equilibrium with its decay or reaction
products: $\alpha + \alpha \rightarrow ^8Be$. The equilibrium concentration
of the $^8Be$ nucleus can be calculated through the Saha equation

$$ N_{12}={N_1 N_2 \over 2} \big({2\pi \over \mu kT}\big)^{3/2} \hbar^3 {(2J +1) \over (2J_1+1) (2J_2 +1)} \rm exp (-{E_R \over k T}) $$
at the relevant temperature $T_6 =11$ and $\rho = 10^5 \; \rm gm \; cm^{-3}$
to be: 

$${N(^8Be) \over N(^4He)} = 5.2 \times 10^{-10} $$
Salpeter suggested that this small quantity of $^8Be$ serves as the seed
for the second stage of the triple $\alpha$-capture into the $^{12}C$ nucleus.
It was however shown by Hoyle \cite{Hoy53} 
that the amount of $^{12}C$ produced
for the conditions inside a star at the tip of the red-giant
branch is insufficient to explain the observed abundance, {\it unless} the
reaction proceeds through a resonance process \cite{Hoy54}. The presence
of such a resonance greatly speeds up the rate of the triple-$\alpha$
process which then proceeds through an s-wave ($l=0$) resonance
in $^{12}C$ near the threshold of $^8Be + \alpha$ reaction.
Since $^8Be$ and $^4He$ both have $J^{\pi} = 0^+$, an s-wave resonance
would imply that the resonant state in question has to be $0^+$ in
the $^{12}C$ nucleus.
Hoyle suggested the excitation energy to be: $E_X \sim 7.68$ MeV
in the $^{12}C$ nucleus and this state was experimentally found by 
W.A. Fowler's group (\cite{Coo57}) with spin-parity:
$J^{\pi} = 0^+$. This state has a total width (\cite{Rol88})
$\Gamma = 8.9 \pm 1.08$ eV, most of which lies in $\Gamma_{\alpha}$,
due to the major propensity of the $^{12}C$ nucleus to break-up through
$\alpha$-decay. (The decay of the excited state of $^{12}C$  by
$\gamma$-rays cannot go directly to the ground state, since the
resonance state as well as the ground state of the $^{12}C$ nucleus
have both $J^{\pi} = 0^+$ and $0^+ \rightarrow 0^+$ decays are
forbidden. This partial width due to gamma-decay is several thousand
times smaller than that due to $\alpha$-decay). 
So, $\Gamma = \Gamma_{\alpha} + \Gamma_{rad} \sim \Gamma_{\alpha}$ and
$\Gamma_{rad} = \Gamma_{\gamma} + \Gamma_{e^+e^-} = 3.67 \pm 0.50$ meV.
Again the radiative width $\Gamma_{rad}$
is dominated by the width due to photon width
deexcitation: $\Gamma_{\gamma} = 3.58 \pm 0.46$ meV. (Note
the scales of {\it milli}electron Volts).
The reaction scheme for the first and the second parts of
the triple-alpha reaction is given in Fig. \ref{fig: 3alpha}.
The locations of the Gamow energy regions near the above resonance state
(for several stellar temperatures) are shown only schematically.

The reaction rate for the $^{12}C$ formation can be calculated
by using the properties of the resonant state and the thermally 
averaged cross-section:

$$r_{3\alpha} = N_{^8Be} N_{\alpha} < \sigma v >_{^8Be + \alpha}$$
Here $N_{^8Be}$ and $N_{\alpha}$ are the number densities
of interacting $^8Be$ and $^4He$ nuclei and the angular brackets
denote thermal averaging over a Maxwell Boltzmann distribution $\psi(E)$.
This averaging leads to:

$$r_{3\alpha} = N_{^8Be} N_{\alpha} \int^{\infty}_0 \psi(E) v(E) \sigma(E) dE$$
with 
$$\psi(E) = {2\over \sqrt \pi} {E\over kT} \rm exp (-E/kT) {dE \over (kTE)^{1/2}}$$
and 
$$\sigma(E) =  \pi \big({\lambda \over 2 \pi}\big)^2 {2J+1 \over (2J_1+1) (2J_2+1)} {\Gamma_1 \Gamma_2 \over (E-E_R)^2 + (\Gamma/2)^2}$$
is the Breit-Wigner resonant reaction cross section with the resonant
energy centroid at $E=E_R$. The total width
$\Gamma$ is a sum of all decay channel widths such as 
$\Gamma_1 = \Gamma_{\alpha}$ and $\Gamma_2 = \Gamma_{\gamma}$.
If the width $\Gamma$ is only a few eVs then the functions $\psi(E)$ and 
$v(E)$ can be pulled out of the integral. Then, the reaction rate
will contain an integral like: $\int_0^{\infty} \sigma_{BW} (E) dE
= 2 \pi (\lambda/2 \pi \hbar)^2 \omega \Gamma_1 \Gamma_2/\Gamma$,
where 
$\omega = (2J+1) / [(2J_1+1) (2J_2+1)]$ 
and the functions pulled out of the integral need to be evaluated at $E= E_R$.
Since  most of the time the excited state of the $^{12}C^*$ breaks-up
into $\alpha$-particles, we have $\Gamma_1 =\Gamma_{\alpha}$ dominating
over $\Gamma_{\gamma}$ and $(\Gamma_1 \Gamma_2 / \Gamma) \sim \Gamma_2$.
This limit usually holds for resonances of energy sufficiently high
so that the incident particle width ($\Gamma_1$) to dominate the natural
width of the state ($\Gamma_2$). In that case, we can use the number
density of the $^8Be$ nuclei in equilibrium with the $\alpha$-particle
nuclei bath as described by Saha equilibrium condition:

$$N(^8Be) = N_{\alpha}^2 \omega f {h^3 \over (2\pi \mu kT)^{3/2}} \rm exp (-E_r/kT)$$
where 
f is the screening factor. 
It is possible to get the overall triple-alpha reaction rate by 
calculating the equilibrium concentration of the excited (resonant) state of
$^{12}C$  reached by the $^8Be + \alpha \rightarrow ^{12}C^*$ reaction
and then multiplying that concentration by the gamma-decay rate 
$\Gamma_{\gamma}/\hbar$ which leads to the final product of $^{12}C$.
So, the reaction rate for the final step of the triple-alpha
reaction turns out to be:

$$r_{3\alpha} = N_{^8Be} N_{\alpha} \hbar^2 \bigg({2\pi \over \mu kT}\bigg)^{3/2} \omega f \Gamma_2 \rm exp (-E_r^{'}/ kT)$$
where $\mu$ is the reduced mass of the reactants $^8Be$ and $\alpha$ particle.
This further reduces by the above argument to:
$$r_{3\alpha \rightarrow ^{12}C} = {N_{\alpha}^3 \over 2} 3^{3/2} \bigg({2\pi \hbar^2 \over M_{\alpha} kT}\bigg)^3 f {\Gamma_{\alpha} \Gamma_{\gamma} \over \Gamma \hbar} \rm exp (-Q/kT)$$
The Q-value of the reaction is the sum of 
$E_R(^8Be + \alpha) = 287 \rm \; keV$ and 
$E_R(\alpha +\alpha) = |Q| = 92 \; \rm keV$ 
and turns out to be: $Q_{3\alpha} = (M_{^{12}C^*} - 3 M_{\alpha})c^2
= 379.38 \pm 0.20 \rm keV$ (\cite{Nol76}). Numerically, the energy generation rate
for the triple-alpha reaction is:

$$\epsilon_{3\alpha} = {r_{3\alpha} Q_{3\alpha} \over \rho} = 3.9 \times 10^{11}{\rho^2 X_{\alpha}^3 \over T_8^3} f \; \rm exp( - 42.94 /T_8) \rm \; erg \; gm^{-1} \; s^{-1}$$
The triple alpha reaction has a very strong temperature dependence: near
a value of temperature $T_0$, one can show that the energy generation rate
is: 

$$\epsilon(T) = \epsilon(T_0) ({T \over T_0})^n$$
where, $n = 42.9/T_8 -3$. Thus at a sufficiently high temperature and density,
the helium gas is very highly explosive, so that a small temperature rise
gives rise to greatly accelerated reaction rate and energy liberation.
When helium thermonuclear burning is ignited in the stellar core under 
degenerate conditions, an unstable and sometimes an explosive condition 
develops. 

\section{Survival of $^{12}C$ in red giant stars and $^{12}C(\alpha, \gamma)^{16}O$ reaction}  

The product of the triple-alpha reactions, $^{12}C$ is burned into $^{16}O$
by $\alpha$-capture reactions:

$$ ^{12}C + \alpha \rightarrow ^{16}O + \gamma $$
If this reaction proceeds too efficiently, then all the carbon will be
burned up to oxygen. Carbon is however the most abundant element in the
universe after hydrogen, helium and oxygen, and the cosmic C/O ratio is 
about 0.6. In fact, the O and C burning reactions  and the conversion of
He into C and O take place in similar stellar core temperature and
density conditions. Major ashes of He burning in Red Giant stars are C
and O. Red Giants are the source of the galactic supply of $^{12}C$ and
$^{16}O$. 
Fortuitous 
circumstances of the energy level structures of these alpha-particle
nuclei are in fact important for the observed abundance of oxygen and
carbon.

\begin{figure}[htb]
\vspace*{-0.5cm}
%                 \insertplot{fig15.eps}
\plotone{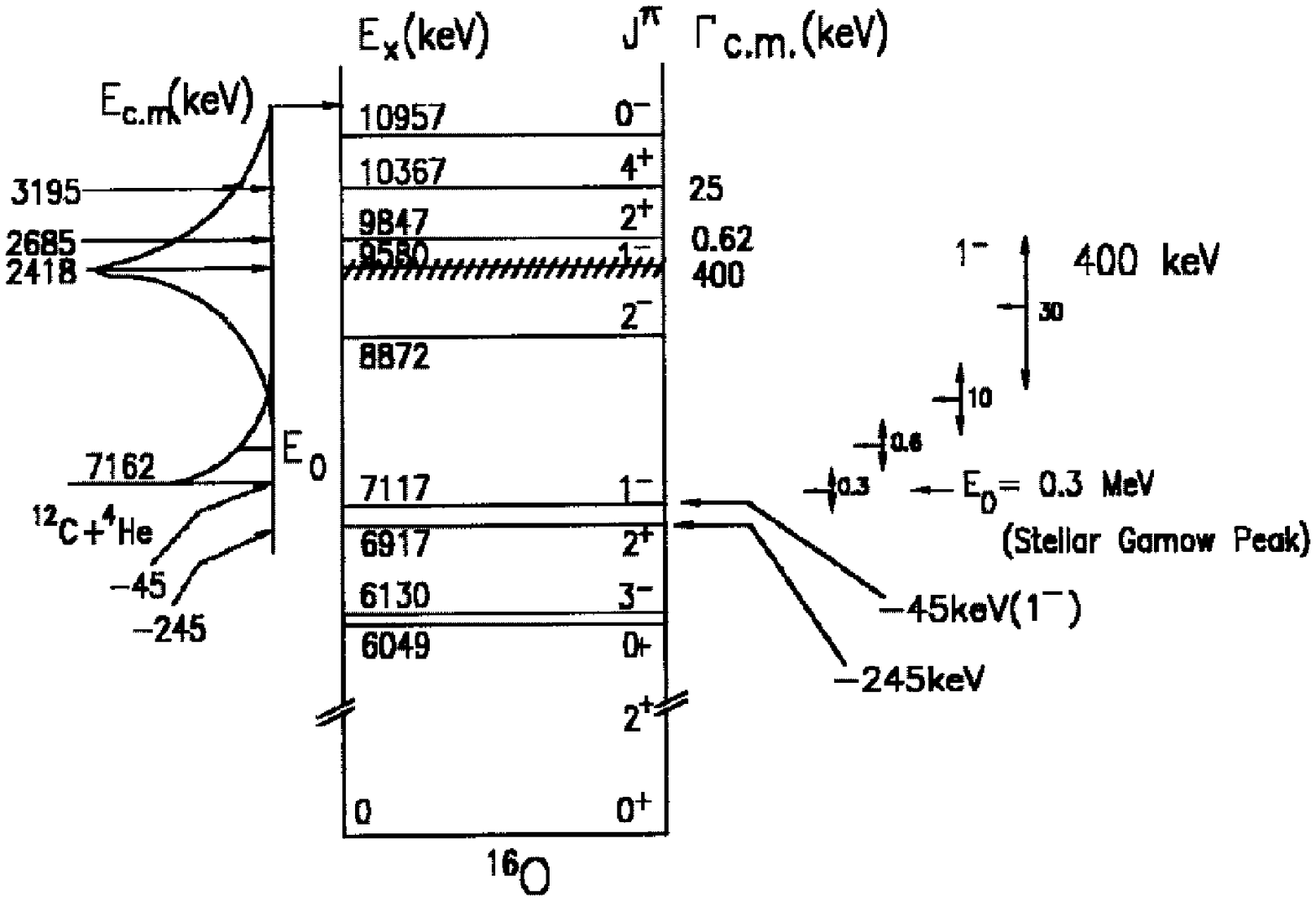}
\vspace*{-0.7cm}
\caption[]{
Energy levels of $^{16}O$ nucleus near and above the
alpha-particle threshold of capture on $^{12}C$. Shown on
the right are effective stellar energy regions corresponding
to the temperatures given near the three-way arrows. The
reaction rate is influenced mainly by the high energy tails
of two subthreshold resonances in $^{16}O$ at $E_R = -45 \; \rm keV$
and $E_R = -245 \; \rm keV$, plus the low energy tail
of another high-lying broad resonance at 9580 keV.
}
\label{fig: 16Olevels}
\end{figure}

For example, if as in the case of the 3$\alpha$ reaction, there was a
resonance in the $^{12}C(\alpha, \gamma)^{16}O$ reaction near the
Gamow window for He burning conditions (i.e. $T_9 \sim 0.1 - 0.2$), then
the conversion of $^{12}C \rightarrow \; ^{16}O$ would have proceeded at a 
very rapid rate.
However, the energy level diagram of $^{16}O$ shows that for temperatures
upto about $T_9 \sim 2$, there is no level available in $^{16}O$ to foster
a resonant reaction behaviour (Fig. \ref{fig: 16Olevels}). But since this
nucleus is found in nature, its production must go through
either: 1) a non-resonant direct capture reaction or 2) nonresonant
captures into the tails of nearby resonances (i.e. subthreshold reactions).
In Fig. \ref{fig: 16Olevels}, also shown on the left of the $^{16}O$
energy levels, is the threshold for the $^{12}C + ^4He$ reaction, drawn at
the appropriate level with respect to the ground state of the $^{16}O$ nucleus.
The Gamow energy regions drawn on the extreme right for temperatures
$T_9 = 0.1$ and above, indicates that for the expected central temperatures,
the effective stellar (centre of mass) energy region is near $E_0 =0.3$ MeV.
This energy region is reached by the low energy tail of a broad resonance
centred at $E_{CM} = 2.42$ MeV above the threshold (the $J^{\pi} = 1^-$ state
at 9.58 MeV above the ground state of $^{16}O$) with a (relatively large)
resonance width of 400 keV. 
On the other hand, there are two subthreshold resonances in $^{16}O$
(at $E_X = 7.12$ MeV and $E_X= 6.92$ MeV), i.e. -45 keV and -245 keV
{\it below} the $\alpha$-particle threshold that have 
$J^{\pi} =1^-$ and $J^{\pi} =2^+$, that contribute to 
stellar burning rate by their high energy tails.
However, electric dipole (E1) $\gamma$-decay of the
7.12 MeV state is inhibited by isospin selection rules.
Had this not been the case, the $^{12}C(\alpha, \gamma)^{16}O$
reaction would have proceeded fast and $^{12}C$ would have been
consumed during helium burning itself.
The two subthreshold states at $-45$ keV and $-245$ keV 
give contributions to the astrophysical S-factor of: 
$S_{1^-}(E_0) = 0.1 \; \rm MeV \; barn$ and $S_{2^+} (E_0) = 0.2 \; \rm MeV \; barn$
respectively at the relevant stellar energy $E_0 = 0.3 \; \rm MeV$.
The state at $E_{CM} = 2.42$ MeV ($J^{\pi} = 1^-$ state at 9.58 MeV)
gives a contribution: $S_{1^-}(E_0) = 1.5 \times 10^{-3} \rm MeV \; barn$.
The total S-factor at $E_0 = 0.3 \; \rm MeV$ is therefore close to
$0.3 \; \rm MeV \; barn$. These then provide low enough S or cross-section
to not burn away the $^{12}C$ entirely to $^{16}O$, so that 
$C/O \sim 0.1 $ at the least.

\begin{figure}[htb]
\vspace*{-0.5cm}
%                 \insertplot{fig16.eps}
\plotone{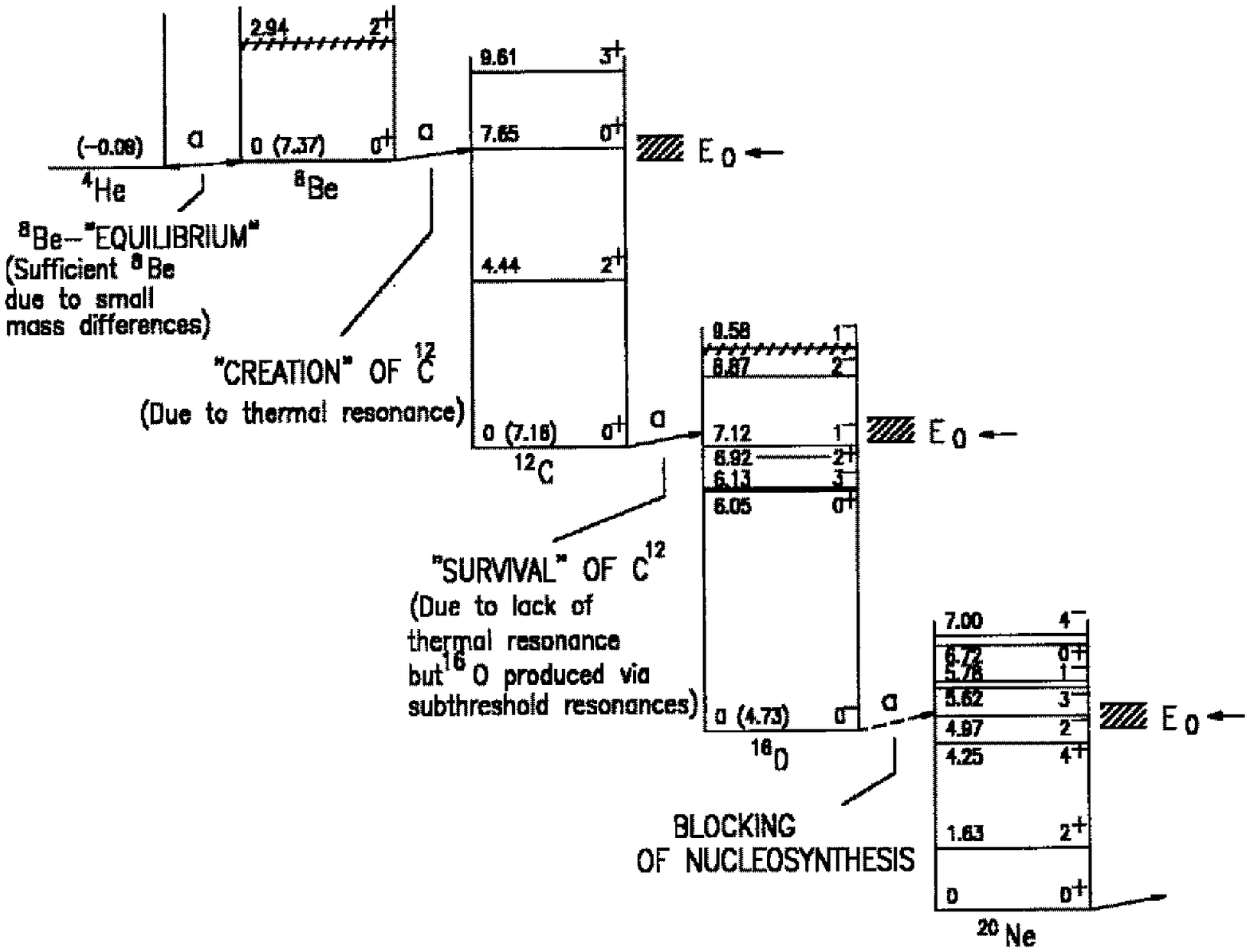}
\vspace*{-0.7cm}
\caption[]{
Energy levels of nuclei participating in thermonuclear reactions
during the helium burning stage in red giant stars (after \cite{Rol88}).
The survival
of both $^{12}C$ and $^{16}O$ in red giants, believed to be the
source of terrestrial abundances depends upon fortuitous
circumstances of nuclear level structures and other properties
in these nuclei.
}
\label{fig: Ne20photo}
\end{figure}

Additionally, $^{16}O$ nuclei are not burnt away by further $\alpha$-capture
in the reaction:

$$^{16}O + ^4He \rightarrow \; ^{20}Ne + \gamma $$
A look at the level schemes of $^{20}Ne$  (see Fig. \ref{fig: Ne20photo})
to
shows the existence of a 
$E_X = 4.97 \; \rm MeV$ state ($J^{\pi} = 2^-$) in the Gamow window.
However, this state cannot form in the resonance reaction
due to considerations of parity conservation (unnatural parity of the
resonant state)\footnote{Whether or not a resonant state can be 
formed or accessed via a given reaction
channel depends upon the angular momentum and parity conservation laws.
The spins of the particles in the entrance channel, $j_1, j_2$ and relative
angular momentum $l$ adds upto the angular momentum of the resonant state
$J = j_1 + j_2 +l$. Therefore, for spinless particles like the closed shell
nuclei $^4He, ^{16}O$ ($j_1 =0, j_2 =0$), we have $J=l$. In the entrance
channel of the  reacting particles, the parity would be:
$(-1)^l \pi(j_1) \pi(j_2) = (-1)^{l=0} (1) (1)$. If the parity of the
resonance state were the same as that of the entrance channel, then
the resultant state would have been a ``natural parity" state. However,
since the 4.97 MeV state in $^{20}Ne$ has an assignment: $J^{\pi} =2^-$,
this is an ``unnatural parity" state.
}. The lower 4.25 MeV state ($J^{\pi} = 4^+$)
in $^{20}Ne$ also cannot act as a subthreshold resonance as it lies too far
below threshold and is formed in the g-wave state. Therefore only
direct capture reactions seem to be operative, which for $(\alpha, \gamma)$
reactions lead to cross-sections in the range of nanobarns or below.
Thus the destruction of $^{16}O$ via: $^{16}O(\alpha, \gamma)^{20}Ne$
reaction proceeds at a very slow rate during the stage of helium
burning in Red Giant stars, for which the major ashes are carbon and oxygen
and these elements have their galactic origin in the Red Giants.

To summarise, the synthesis of 
two important elements for the evolution of life as
we know on the earth have depended upon fortuitous circumstances of
nuclear properties and selection rules for nuclear reactions.
These are: 1) the mass of the unstable lowest (ground) state of $^8Be$
being close to the combined mass of two $\alpha$-particles; 2) there is 
a resonance in $^{12}C$ at 7.65 MeV which enhances the alpha addition
reaction (the second step); and 3) parity conservation has protected $^{16}O$
from being destroyed in the $^{16}O (\alpha, \gamma)^{20}Ne$ reactions
by making the 4.97 MeV excited state in $^{20}Ne$ of unnatural parity.

\section{Advanced stages of thermonuclear burning}  

As the helium burning progresses, the stellar core is increasingly
made up of C and O. At the end of helium burning, all hydrogen and helium
is converted into a mixture\footnote{Note however the caveat: if the amount of $^{12}C$ is little (either due to
a long stellar lifetime of He burning or due to a larger rate of the
$^{12}C + \alpha \rightarrow ^{16}O + \gamma$ reaction whose estimate
outlined in the earlier section is somewhat uncertain), then the star may
directly go from He-burning stage to the O-burning or Ne-burning stage
skipping C-burning altogether (\cite{Woo86}).
} of C and O, and since H, He  
are most abundant elements in the original gas from which the
star formed, the amount of C and O are far more than the traces
of heavy elements in the gas cloud. 
Between these two products, the Coulomb barrier
for further thermonuclear reaction involving the products is lower
for C nuclei. At first the C+O rich core is surrounded by He burning shells
and a helium rich layer, which in turn may be surrounded by hydrogen
burning shell and the unignited hydrogen rich envelope.
When the helium burning ceases to provide sufficient power, the star begins
to contract again under its own gravity and as implied by the Virial
theorem the temperature of the helium exhausted core rises. The contraction
continues until either the next nuclear fuel begins to burn at rapid
enough rate or until electron degeneracy pressure halts the infall.

\subsection{Carbon burning}

\begin{figure}[htb]
\vspace*{-0.5cm}
%                 \insertplot{fig17.eps}
\plotone{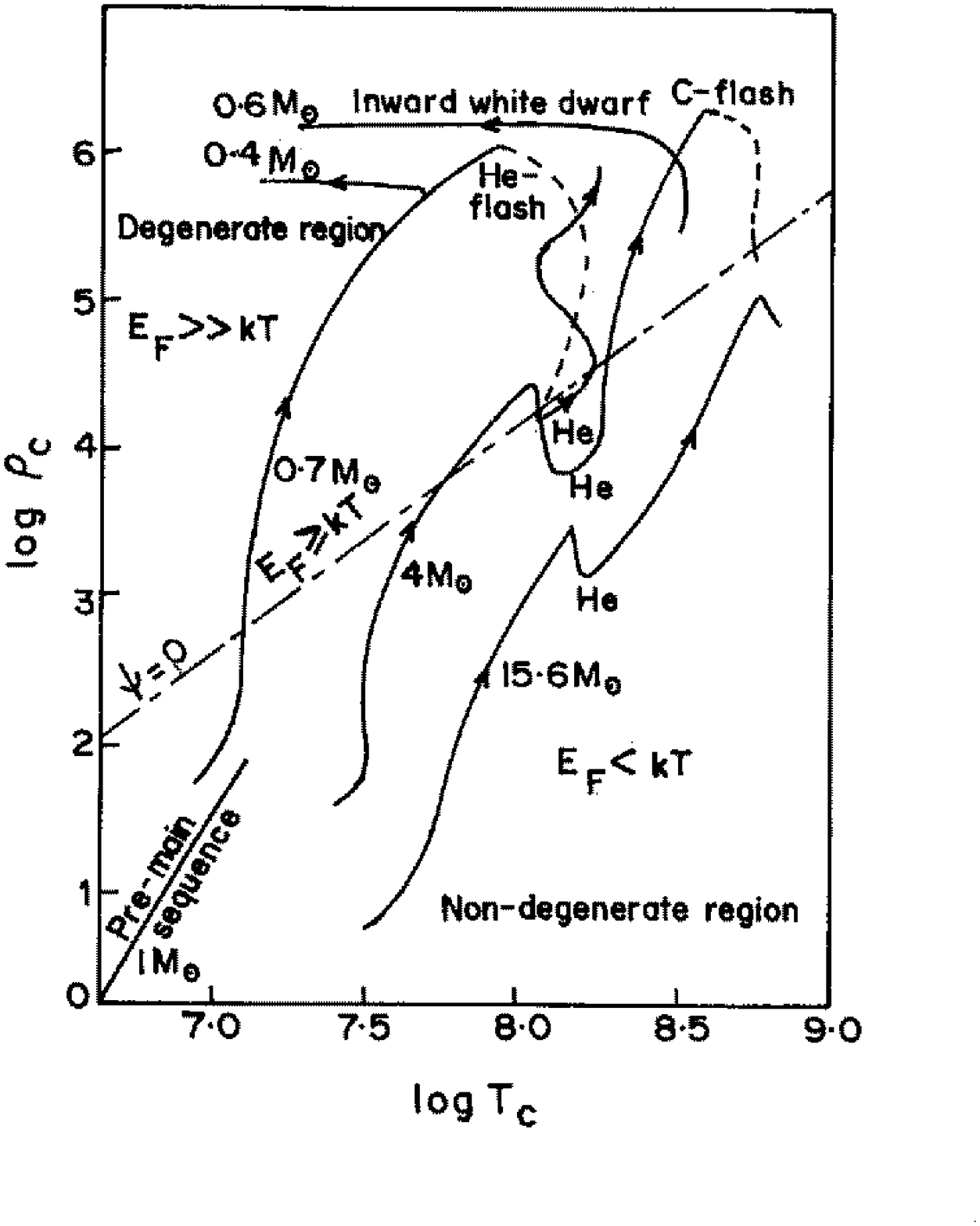}
\vspace*{-0.7cm}
\caption[]{
Tracks in the core temperature, density plane of 
stars of various masses (at the start of hydrogen burning
i.e. main sequence masses). Note that a star of mass
$M \sim 15 M_{\odot}$ ignites all its fuels in 
non-degenerate conditions, whereas a star of mass 
$M \sim 4 M_{\odot}$ ignites carbon under strongly
degenerate conditions. (After \cite{Hay62}).
}
\label{fig: Hayashi}
\end{figure}

Stars somewhat more massive than about $0.7 \; \rm M_{\odot}$ contract
until the temperature is large enough for carbon to interact with itself
(stars less massive may settle as degenerate helium white dwarfs). For
stars which are more massive than $M \geq 8-10 \; M_{\odot}$ (mass on the
main sequence, - {\it not} the mass of the C+O core), the contracting C+O
core remains nondegenerate until C starts burning at 
$T \sim 5 \times 10^8 K$ and $\rho = 3 \times 10^6 \; \rm g cm^{-3}$. 
Thereafter sufficient power is 
generated and the contraction stops and quiescent (hydrostatic, not
explosive) C-burning proceeds (see Fig. \ref{fig: Hayashi}). 

The combined mass of two reacting $^{12}C$ nuclei falls at an excitation
energy of 14 MeV in the compound nucleus of $^{24}Mg$. At this energy
there are many compound nuclear states, and the most effective range
of stellar energies (the Gamow window) at the relevant temperature is
about 1 MeV; hence a number of resonant states can contribute to the
decay of the compound nucleus, and even the large angular momentum resonances
may be important because the penetration factors in massive nuclei are not
affected by centrifugal barriers.
The carbon on carbon burning can proceed through multiple, 
energetically allowed reaction channels, listed 
%in Table \cite{tab: C12C12}.
below:
$$ ^{12}C + ^{12}C  \rightarrow \;^{20}Ne + ^4He \; \; (Q = 4.62 \; \rm MeV) \\$$
$$ \newline \; \; \; \; \; \; \; \; \; \;    \rightarrow \; ^{23}Na + p \; \; (Q = 2.24 \; \rm MeV) \\$$
$$ \newline \; \; \; \; \; \; \; \; \; \; \;    \rightarrow \; ^{23}Mg + n \; \; (Q = -2.62 \; \rm MeV) \\$$
At the temperatures where carbon burning starts, the neutron liberating
reactions requires too much particle kinetic energy to be effective.
In addition, based on laboratory measurements at higher energies
compared to the stellar energies,
the electromagnetic decay channel ($^{24}Mg + \gamma$) and the  
three particle channel ($^{16}O + 2\alpha$) have lower probability compared
to the two particle channels: $^{23}Na + p$ and $^{20}Ne + \alpha$.
The latter two channels have nearly equal probabilities (see 
\cite{Cla68};
at the lowest centre of mass energies for which cross-sections
are measured in the laboratory for the proton and $\alpha$ channels,
(i.e. about 2.45 MeV \cite{Maz72}), the branching ratios were
$b_p \sim 0.6$ and $b_{\alpha} \sim 0.4$),
and therefore the direct products of carbon burning
are likely to be $^{23}Na$, $^{20}Ne$, protons and alpha particles.
The rate for this reaction per pair of $^{12}C$ nuclei is (\cite{Ree59}):

$$\rm log \lambda_{12,12} = log f_{12,12} + 4.3 - {36.55(1 + 0.1T_9)^{1/3}\over T_9^{1/3} } -{2\over3} log T_9 $$
the factor $f_{12,12}$ is a screening factor. Now, at the temperatures of
$^{12}C$ burning, the liberated protons and alpha particles can be quickly
consumed through the reaction chain: $^{12}C(p, \gamma)^{13}N(e^+\nu_e)^{13}C
(\alpha, n)^{16}O$. Thus, the net effect is that
the free proton is converted into a free neutron (which may be further 
captured) and the $\alpha$-particle is consumed with $^{12}C$ into $^{16}O$. 
The $\alpha$-particles are also captured by other alpha-particle nuclei,
resulting in, at the end of carbon burning in nuclei like: 
$^{16}O$, $^{20}Ne$, $^{24}Mg$ and $^{28}Si$. These secondary reactions augment
the energy released by the initial carbon reaction and Reeves (1959) estimated
that each pair of $^{12}C$ nuclei release about 13 MeV of energy.
Towards the end of carbon burning phase there are also other reactions
such as: $^{12}C + ^{16}O$ and $^{12}C + ^{20}Ne$ which take place.
But these are less rapid and are not expected to play major roles compared
to the $^{12}C+^{12}C$ reactions, due to their increased Coulomb barriers.
A recent discussion of the heavy ion reactions involving C and
O is contained in \cite{Arn96} section 3.6>

During the carbon-burning and subsequent stages, the dominant energy
loss from the star is due to neutrinos streaming out directly
from the stellar thermonuclear furnace, rather than by photons from the
surface. The neutrino luminosity is a sensitive function of core
temperature and quickly outshines the surface photon luminosity of
the star at carbon burning stage. The (thermal) evolutionary timescale
of the star, due to the neutrino emission becomes very short and the core
evolves rapidly, -- so rapidly (compared to the ``cooling" timescale
Kelvin-Helmholtz time: $\tau_{KH} \sim GM^2/R L_{ph}$) that the conditions
in the core are ``not communicated" to the surface, since this communication
happens by photon diffusion. The surface conditions (e.g. the temperature)
of the star then does not markedly evolve as the core goes beyond
the carbon burning stage, and it may not be possible just by looking
at a star's surface conditions whether the core is close to a
supernova stage or has many thousands of years of hydrostatic thermonuclear
burning to go.

\subsection{Neon burning}

The result of carbon burning is mainly neon, sodium and magnesium,
but aluminium and silicon are also produced in small quantities
by the capture of $\alpha$, p and n released during carbon burning.
When carbon fuel is exhausted, again the core contracts and its temperature
$T_c$ goes up. At approximately $T_9 \sim 1$, energetic photons from
the high energy tail of the Planck distribution function can begin to
disintegrate the $^{20}Ne$ ash (see Fig. \ref{fig: Ne20photo})
so that one has the reaction: $^{20}Ne + \gamma \rightarrow ^{16}O +^4He$.

Nucleons in a nucleus are bound with typical binding energy of several to $~8$ 
MeV.  One would require an energetic
$\gamma$-ray photon to photo-eject
a nucleon. Two nucleon ejection would require more energy. Alpha particles
are however released at approximately the same energy as a nucleon due to
the low separation energy of an alpha particle in the nucleus. For example,
the alpha separation energy in $^{20}Ne$ is 4.73 MeV. Thus, the major 
photonuclear reactions are: $(\gamma, n), (\gamma, p)$ and $(\gamma, \alpha)$
processes. For a photodisintegration reaction to proceed through an excited
state $E_X$ in the mother, the decay rate is:-

$$\lambda(\gamma, \alpha) = \bigg[\rm exp \big(-{E_X \over kT}\big) {2J_R +1 \over 2J_0 + 1} {\Gamma_{\gamma} \over \Gamma} \bigg] \times {\Gamma_{\alpha} \over \hbar} $$
In the above equation, the first factor in square brackets on the RHS is
the probability of finding the nucleus in the excited state $E_X$ and
spin $J_R$ (with $J_0$ being the ground state spin), while the second factor
$\Gamma_{\alpha} / \hbar$ is the decay rate of the excited state with an
alpha particle emission. Now since $E_X = E_R + Q$, we have:

$$\lambda(\gamma, \alpha) = {\rm exp(- Q/kT) \over \hbar (2J_0 +1)} (2J_R +1) {\Gamma_{\alpha} \Gamma_{\gamma} \over \Gamma} \rm exp(-E_R / kT) $$
At $T_9 \geq 1$, the photodisintegration is dominated by the 5.63 MeV
level in $^{20}Ne$ (see Fig. \ref{fig: Ne20photo}). At approximately
$T_9 \sim 1.5$, the photodissociation rate becomes greater than the rate
for alpha capture on $^{16}O$ to produce $^{20}Ne$ (i.e. the reverse
reaction), thus leading effectively to the net dissociation of $^{20}Ne$.
The released $^4He$ reacts with the unspent $^{20}Ne$ and leads to:
$^{4}He + \; ^{20}Ne \rightarrow \; ^{24}Mg + \gamma$. Thus the net result of
the photodissociation of two $^{20}Ne$ nuclei is: 
$2 \times ^{20}Ne \rightarrow \; ^{16}O + ^{24}Mg$ 
with a net Q-value of 4.58 MeV. The brief neon
burning phase concludes at $T_9$ close to $\sim 1$.

\subsection{Oxygen burning}

At the end of the neon burning the core is left with a mixture of
alpha particle nuclei: $^{16}O$ and $^{24}Mg$. After this another core
contraction phase ensues and the core heats up, until at $T_9 \sim 2$,
$^{16}O$ begins to react with itself:

$$ ^{16}O + ^{16}O  \rightarrow ^{28}Si + ^4He \\$$ 
$$ \newline \; \; \; \; \; \; \; \; \; \;    \rightarrow ^{32}S + \gamma$$
The first reaction takes place approximately $45 \%$  of the time
with a Q-value of 9.593 MeV. In addition to Si and S, the oxygen
burning phase also produces, Ar, Ca and trace amounts of Cl, K, etc
upto Sc. Then at $T_9 \sim 3$, the produced $^{28}Si$ begins to
burn in what is known as the Si burning phase.

\subsection{Silicon burning}

As we have seen, most of the stages of stellar burning involve
thermonuclear fusion of nuclei to produce higher Z and A nuclei.
The first exception to this is neon burning where the photon field is
sufficiently energetic to photodissociate neon, before the temperature
rises sufficiently to allow fusion reactions among oxygen nuclei
to overcome their Coulomb repulsion. Processing in the neon burning
phase takes place with the addition of helium nuclei to the undissociated
neon rather than overcoming the Coulomb barrier of two neon nuclei. 
This trend continues in the silicon burning phase. In general, a 
photodisintegration channel becomes important when the temperature
rises to the point that the Q-value, i.e. the energy difference
between the fuel and the products is smaller than approximately
$30 k_BT$ (\cite{Hix96}).

With typical Q-values for reactions among stable nuclei above
silicon being 8-12 MeV, photodisintegration of the nuclear products of
neon and oxygen burning begins to play an important role once
the temperature exceeds: $T_9 \geq 3$. Then nuclei with smaller binding 
energies are destroyed by photodissociation
in favour of their more more tightly bound
neighbours, and many nuclear reactions involving $\alpha$-particles,
protons and neutrons interacting with all the nuclei in the mass range
$A= 28-65$ take place. 
In contrast to the previous burning stages where only a few
nuclei underwent thermonuclear reactions upon themselves, here the nuclear
reactions are primarily of a rearrangement type, in which a particle is
photoejected from one nucleus and captured by another and a given fuel
nucleus is linked to a product nucleus by a multitude of reaction chains
and cycles and it is necessary to keep track of many more
nuclei (and many reaction processes involving these) than for previous
burning stages..
More and more stable forms of the nuclei form in a nuclear reaction
network as the rearrangement proceeds. Since there exists a maximum in the
binding energy per nucleon at the $^{56}Fe$ nucleus, the rearrangements
lead to nuclei in the vicinity of this nucleus (i.e. iron-group nuclei).

In the mass range $A= 28-65$, the levels in the compound nuclei that form
in the reactions during silicon burning are so dense that they overlap.
Moreover, at the high temperatures that are involved ($T_9 = 3-5$),
the net reaction flux may be small compared to the large forward
and backward reactions involving a particular nucleus and a quasi-equilibrium
may ensue between groups of nuclei which are connected between separate
groups by a few, slow, rate-limiting reactions (``bottlenecks"). However,
as the available nuclear fuel(s) are consumed and thermal energy is removed
due to escaping neutrinos, various nuclear reactions may no longer occur
substantially rapidly (``freeze-out"). Thielemann and Arnett (\cite{Thi85})
found that for cores of massive stars in hydrostatic cases, the bottlenecks
between quasi-equilibrium (QSE) groups coincided with Z=21 nuclei
whereas for lower mass stars, lower temperatures and $Y_e$ and higher density
this bridge involved neutron rich isotopes of Ca capturing protons.
Hix and Thielemann \cite{Hix96} 
discussed and contrasted these results with those
of earlier workers and in general the reaction flow across the boundary of
the QSE  groups are influenced by the neutronisation of the material,
i.e. the overall $Y_e$. It is in this context that weak interaction processes
such as electron capture and beta decay of nuclei are important, by
influencing the $Y_e$ and thereby the reaction flow. These ultimately
affect both the stellar core density and entropy structures, and it is
important to track and include the changing $Y_e$ of the core material
not only in the silicon burning phase, but even from earlier oxygen burning
phases. The calculation of stellar weak processes on nuclei has spawned
extensive literature (see \cite{Ful82}, \cite{Kar98} etc., and \cite{Lan02}
for a review).

In summary, a few key points concerning the thermonuclear burning of $^{28}Si$
are as follows:-

$\bullet$ Direct thermonuclear fusion of two $^{28}Si$ nuclei does not take place
because their Coulomb barrier is too high. Instead thermonuclear fusion takes
place by successive additions of $\alpha$-particles, neutrons and protons.

$\bullet$ Although this is actually a large network of nuclear reactions
it is called ``silicon burning" because $^{28}Si$ offers the largest resistance
to photo-dissociation because of its highest binding energy among intermediate
mass nuclei.

$\bullet$ The source of the $\alpha$-particles which are captured by $^{28}Si$
and higher nuclei is $^{28}Si$ itself. Silicon, sulphur etc. partially
melt-down into $\alpha$-particles, neutrons and protons by photo-dissociation.
These then participate in reaction networks involving quasi-equilibrium clusters
linked by ``bottleneck" links.

$\bullet$ Although beta decay and electron captures on stellar
core nuclei do not produce
energy in major ways they nevertheless play a crucial role in shifting the
pathways of nuclear and thermodynamic evolution in the core conditions.
These ultimately determine the mass of the core and its entropy
structure which finally collapses in a supernova explosion.

\section{Nucleosynthesis beyond Iron: neutron induced reactions}

So far we have been dealing primarily with charged particle reactions and
photodisintegration which lead to the production of lighter elements
($1\leq A \leq 40$) and the recombination reactions for the production of
elements $40 \leq A \leq 65$.
However, the heavier elements ($A\geq 65$), because of their high charge and
relatively weak stability, cannot be produced by these two processes.
It was therefore natural to investigate the hypothesis of neutron induced
reactions on the elements that are formed already in the various
thermonuclear burning stages, and in particular on the iron-group elements.
The study of the nuclear reaction chains in stellar evolution shows
that during certain phases large neutron fluxes are released in the core
of a star. On the other hand, the analysis of the relative abundance
of elements shows certain patterns which can be explained in terms of
the neutron absorption cross-sections of these elements. If the heavier
elements above the iron peak were to be synthesised during for example
in charged particle thermonuclear reactions during
silicon burning, their abundance would drop very much more steeply
with increasing mass (larger and larger Coulomb barriers) than the observed
behaviour of abundance curves which shows a much lesser than expected decrease.
Based upon the abundance data of Suess and Urey (\cite{Sue56}),
Burbidge et al (\cite{Bur57} hereafter $\rm B^2FH$) and independently Cameron
(\cite{Cam57}) argued that heavy elements are made instead by thermal neutron
capture. 

These authors realised that two distinct neutron processes
are required to make the heavier elements. The slow neutron capture process
(s-process) has the lifetime for $\beta$-decay $\tau_{\beta}$ shorter than
the competing neutron capture time $\tau_n$ (i.e. $\tau_{\beta} \leq \tau_n$). 
This makes the s-process nucleosynthesis run through the valley of 
$\beta$-stability. The rapid neutron capture r-process on the other hand 
requires $\tau_n \ll \tau_{\beta}$. This process takes place in extremely
neutron-rich environments, for the neutron capture timescale is
inversely proportional to the ambient neutron density. The r-process,
in contrast to the s-process, goes through very neutron rich and unstable
nuclei that are far off the valley of stability. The relevant properties
of such nuclei are most often not known experimentally, and are usually
estimated theoretically. Some of the key parameters are the half-lives of the
$\beta$-unstable nuclei along the s-process path. But the nuclear half-life
in stellar environment can change due to transitions from not just the
ground state of the parent nucleus, but also because its excited states
are thermally populated. In the r-process, the $\beta$-decay properties
of the nuclei regulate the reaction flow to larger charge numbers
and determine the resultant abundance pattern and the duration of the
process. The r-process lasts for typically few seconds, in an intense
neutron density environment: $n_n \sim 10^{20} - 10^{25} \rm cm^{-3}$.
In comparison, the neutron densities in the s-process are much more
modest, say: $n_n \sim 10^8 \rm cm^{-3}$; these neutron irradiation can
take place for example in the helium burning phase of Red Giant stars.
Nuclei above the iron group up to about A = 90 are produced in massive stars
mainly by the s-process. Above A = 100 the s-process does very little in
massive stars, although there are
redistributions of some of the heavy nuclei. 
Most of the s-process above mass 90 is believed
to come from Asymptotic Giant Branch stars. For a recent discussion of 
nucleosynthesis in massive stars, see Rauscher et al {\cite{Rau99}.

\section{Conclusions}

The nuclear reactions in the various stages of hydrostatic thermonuclear
burning of massive stars have been discussed in this article. We discussed
mainly the charged particle reactions, but also briefly mentioned 
the neutron induced
reactions and photonuclear processes. What was discussed in the lectures,
but could not be included in these notes was a description of the subsequent
stage, i.e. the gravitational collapse of the core of the massive star
under its own gravity that leads to a supernova explosion. These are
extremely energetic explosions where the observable energy in the kinetic
energy of the exploded debris and electromagnetic radiation add up to several
times $10^{51} \; \rm erg$. The actual energy scale is typically
$3\times 10^{53} \; \rm erg$ or higher, but most of this is radiated away in 
neutrinos. Although the full understanding of the process of explosion
in a gravitational collapse leading to a supernova has not been achieved
despite several decades of theoretical and computational work, a watershed
in the field was achieved observationally
when a supernova exploded close by in a satellite galaxy
of our own, namely SN1987A in the Large Magellanic Cloud (LMC). A few neutrinos
were detected from this supernova, which were the first detections 
of the astrophysical neutrinos from outside of our solar system. 
By using the energetics of the neutrinos, their approximate
spectral distribution, the distance to the LMC it was possible to show
that the overall energy of the explosion was indeed 
$E_T \sim 2-3 \times 10^{53} \rm erg$.
In addition, the duration of the neutrino burst was of the order of a few 
seconds as opposed to a few milliseconds, signifying that the observed
neutrinos had to diffuse out of the dense and opaque stellar matter
as predicted theoretically, instead of directly streaming out of the core.
The spectral characteristics also indicated that the object that is
radiating the neutrinos was indeed very compact, roughly of the same
dimensions as that of a protoneutron star. Thus  SN1987A
provided the observational confirmation of the broad aspects of
the theoretical investigation of stellar collapse and explosion.
For a review of the understanding of the 
astrophysics of SN1987A, see \cite{Arn89}. 

Physicists are now gearing up to detect not only another supernova
in our own galaxy, but by hoping to build very large neutrino detectors, they
aim to detect supernova neutrinos from the local group of galaxies
(\cite{Cas00}, \cite{Cha00}, \cite{Mur00}).
As neutrinos from the supernova travel directly out from the core,
they arrive a few hours ahead of the light flash from the exploding
star, since electromagnetic radiation can only be radiated from
the surface of the star, and it takes the supernova shock launched at
the deep core several hours to travel to the surface. In the case
of SN1987A, this time delay was useful in estimating the size of the
star that exploded and was consistent with other (optical) spectroscopic data
in this regard. Thus some advance warning ahead of the optical brightening
of a supernova can be gotten from a ``neutrino watch".
In AD 1604 when excitement 
arose over the discovery of what is to be later known as ``Kepler's supernova",
Galileo was criticised by the
Padua city council, for not having discovered it.
Galileo apparently replied that he had more important things to
do than to gaze out of the window,
on the slim chance that he might catch something unusual
(subsequently however, he participated in the lively discussions
that took place about this new object in the sky). Physicists
however would now be able to
give an advance warning of an impending galactic supernova by a
worldwide array of neutrino detectors connected loosely through
the internet
(SN Early Warning System or SNEWS\footnote{See the site: 
http://hep.bu.edu/$\sim$snnet/}) which will notify
astronomers to turn their optical, UV and other
telescopes to the right direction
when they find a burst of neutrinos characteristic
of a supernova explosion. This advance warning will be of
importance to catch the characteristics of the early ultraviolet
and soft x-ray emission from the exploding star, in turn giving the
structure of the outer layers of the progenitor star.
Crucial input to the field of nuclear astrophysics is
also coming from laboratory experiments involving
radioactive ion beams (RIB). Short lived nuclei can only
be studied close to their sites of formation before
they decay away. Such complex facilities  will further 
define through experimental input, the
future of nuclear astrophysics.

\section*{Acknowledgments}
I thank the organisers of the SERC School on Nuclear Physics, in particular,
Prof. I. M. Govil for the invitation
to visit Panjab University, Chandigarh where these lectures were
given. I also thank him and Dr. R. K. Puri for
their patience for this manuscript and
Dr. B. K. Jain for his interest in
the interface of 
nuclear physics and astrophysics.
Dr. V. Nanal and P. Chandra are thanked
for reading and commenting
on the lecture notes. 
Research in nuclear astrophysics
at Tata Institute is a part of the Plan Project: 10P-201.

\vfill\eject

\begin{thebibliography}{99}  

\bibitem{Arn96}1. W. D. Arnett, {\it Supernovae and Nucleosynthesis}, Princeton
University Press (1996).

\bibitem{Arn89}2. W. D. Arnett, J. N. Bahcall,  R. P. Kirshner, S. E. Woosley 
{\it Ann. Rev. Astron. Astrophys.}, 27 (1989), 629. 

\bibitem{Atk29}3. R. d'E. Atkinson, and F. G. Houtermans {\it Z. Phys.} 
{\bf 54} (1929) 656.

\bibitem{Bah03}4. J. N. Bahcall, M. C. Gonzalez-Garcia, C. Pefia-Garay 
{\it Phys. Rev. Lett.} {\bf 90}, (2003) 131301.

\bibitem{Bah02}5. J. N. Bahcall, L. S. Brow,  A. Gruzinov and R. F. Sawyer {\it A.\& A.} {\bf 383}, (2002) 291.

\bibitem{Bah89}6. J. N. Bahcall, {\it Neutrino Astrophysics}, Chapter 3, 
Cambridge University Press (1989); see also Bahcall's web pages at:
http://www.sns.ias.edu/$\sim$jnb/.

\bibitem{Bah69}7. J. N. Bahcall and R. M. May {\it Ap. J.} {\bf 155}, (1969) 501.

\bibitem{Bah69b}8. J. N. Bahcall and C. P. Moeller {\it Ap. J.} {\bf 155}, (1969) 511.

\bibitem{Bet36}9. H. A. Bethe and R.F. Bacher {\it Rev. Mod. Phys.} {\bf 8}, (1936) 82.

\bibitem{Bet37}10. H. A. Bethe {\it Rev. Mod. Phys.} {\bf 9}, (1937) 69.

\bibitem{Bet38}11. H. A. Bethe, and C. L. Critchfield {\it Phys. Rev.} {\bf 54}
(1938) 248.

\bibitem{BetCri38}12. H. A. Bethe, and C. L. Critchfield {\it Phys. Rev.} {\bf 54}
(1938) 862. 

\bibitem{Bet39}13. H. A. Bethe {\it Phys. Rev.} {\bf 55}, (1939) 103 and 434.

\bibitem{Bur57}14. E. M. Burbidge, G. R. Burbidge, W. A. Fowler and F. Hoyle {\it Rev. Mod. Phys.} {\bf 29}, (1957) 547.

\bibitem{Cam57}15. A. G. W. Cameron {\it Stellar Evolution, Nuclear Astrophysics and Nucleogenesis} (1957) Chalk River Report CRL-41. 

\bibitem{Car94}16. B. J. Carr {\it Ann Rev Astr. Astrophys.} {\bf 419} (1994) 904. 

\bibitem{Car84}17. B. J. Carr, J. Bond, and W. D. Arnett, {\it Ap. J.} 
{\bf 277} (1984) 445.

\bibitem{Cas00}18. D. Casper {\it UNO A Next Generation Detector for Nucleon 
Decay and Neutrino Physics} 
$\rm meco.ps.uci.edu/lepton\_workshop/talks/casper/uno.pdf$
(2000). 

\bibitem{Cha00}19. K. J. Chang {\it hep-ex/0005046} (2000). 

\bibitem{Chr02}20. N. Chriestlieb et al. {\it Nature} {\bf 419} (2002) 904. 

\bibitem{Cla68}21. D. D. Clayton, {\it Principles of Stellar Evolution and
Nucleosynthesis}, University of Chicago Press (1984).

\bibitem{Coo57}22. C. W. Cook, W. A. Fowler, C. C. Lauritsen, and T. Lauritsen, 
{\it Phys. Rev.} {\bf 107} (1957) 508. 

\bibitem{Dav68}23. R. Davis, D. S. Harmer, and K. C. Hoffman {\it Phys.
Rev. Lett.} {\bf 20} (1968) 1205.

\bibitem{Edd20}24. A. S. Eddington, {\it Observatory}, {\bf 43}, (1920), 341.

\bibitem{Fer51}25. E. Fermi, {\it Nuclear Physics, Course Notes}, University of
Chicago Press (1951), p.83.

\bibitem{Fil83}26. B. W. Filippone, et al., {\it Phys Rev} {\bf C28}, (1983), 2222.

\bibitem{Fow77}27. W. A. Fowler {\it Proc. Welch Found. Conf. on Chemical Research},
ed. W.D. Milligan (Houston Univ. Press) (1977), p. 61.

\bibitem{Fri51}28. E. Frieman and L. Motz, {\it Phys Rev} {\bf 89}, (1951), 648.

\bibitem{Ful82}29. G. M. Fuller, W. A. Fowler and M. J. Newman, {\it Ap. J.}, {\bf252} (1982), 715.

\bibitem{Gui94}30. M. Guidry (1994) http://csep10.phys.utk.edu/guidry/RIB-7-94html/cno.html.

\bibitem{Hax99}31. W. Haxton, {\it Nuclear Astrophysics Course} 
http://ewiserver.npl.washington.edu/phys554/phys554.html (1999).

\bibitem{Hay62}32. C. Hayashi, R. Hoshi and D. Sugimoto, {\it Progr. Theor. Phys. Suppl.} {\bf 22}, (1951).

\bibitem{Hir87}33. K. Hirata, T. Kajita, M. Koshiba, et al. {\it Physical Rev. 
Lett.}, {\bf 58}, (1987) 1490. 

\bibitem{Hix96}34. W. R. Hix and F. K. Thielemann {\it Ap. J.} {\bf 460}, (1996), 869.

\bibitem{Hoy54}35. F. Hoyle, {\it Ap. J. Sup.} {\bf 1}, (1954), 121.

\bibitem{Hoy53}36. F. Hoyle, D. N. F. Dunbar, W. A. Wenzel, and W. Whaling
{\it Phys. Rev.} {\bf 92}, (1953), 1095.

\bibitem{Kar98}37.  K. Kar, S. Chakravarti, A. Ray and S. Sarkar {\it J. Phys. G.} {\bf 24}, (1998), 1641.

\bibitem{Kir78}38. T. Kirsten, {\it The origin of the solar system}, ed. S. F. Dermott (New York: Wiley) (1978), p. 267.

\bibitem{Lan02}39. K. Langanke and G. Martinez-Pinedo {\it nucl-th/0203071}, Rev. Mod. Phys.{\bf 75}, (2003), 819.

\bibitem{Maz72}40. M. Mazarkis and W. Stephens, {\it Ap. J.} {\bf 171}, (1972) L97.

\bibitem{Mur00}41. M. V. N. Murthy et al., {\it Pramana} {\bf 54}, (2002) 347.

\bibitem{Nol76}42. J. A. Nolen and S. M. Austin, {\it Phys Rev}, C13, (1976), 1773.

\bibitem{Opi51}43. E. J. \"Opik, {\it Proc. Roy. Irish Acad.}, A54, (1951), 49.

\bibitem{Ornl99}44. Oak Ridge National Lab, {\it Nuclear Data for Nuclear
Astrophysics} 
http://www.phy.ornl.gov/astrophysics/data

\bibitem{Lbl98}45. Lawrence Berkeley National Lab, {\it Nuclear
Astrophysics Reaction Rates} 
http://ie.lbl.gov/astro/astrorate.html

\bibitem{Rau99}46. T. Rauscher, A. Heger, R. D. Hoffman, and S. E. Woosley
{\it Ap. J.}, 576, (1999)  323.

\bibitem{Ree68}47. H. Reeves, {\it Stellar Evolution and Nucleosynthesis},
Gordon and Breach, New York (1968).
 
\bibitem{Ree59}48. H. Reeves and E. E. Salpeter, {\it Phys. Rev.}, 116, (1959) 1505.

\bibitem{Rol88}49. C. E. Rolfs and W. S. Rodney, {\it Cauldrons in the Cosmos},
University of Chicago Press (1988).

\bibitem{Rut29}50. E. Rutherford, {\it Nature}, 123, (1929) 313.

\bibitem{Sch83}51. G. K. Schenter and P. Vogel, {\it Nucl. Sci. Engg.},
83, (1983) 393.

\bibitem{Sal57}52. E. E. Salpeter, {\it Phys. Rev.}, 107, (1957) 516.

\bibitem{Sal54}53. E. E. Salpeter, {\it Australian J. Phys}, 7, (1954) 373.

\bibitem{Sal52}54. E. E. Salpeter, {\it Phys. Rev.}, 88, (1952) 547; also
{\it Ap. J.}, 115, (1952), 326.

\bibitem{Sal03}55. R. Salvaterra and A. Ferrara, {\it Mon. Not. Roy. Astr. Soc}, 340, (2003) L17.

\bibitem{Sue56}56. H. E. Suess and H. C. Urey, {\it Rev. Mod. Phys.} 28, (1956)  53.

\bibitem{Thi85}57. F. K. Thielemann and W. D. Arnett {\it Ap. J.}, 295, (1985) 264.

\bibitem{Wal81}58. R. K. Wallace and S. E. Woosley, {\it Ap.J. Sup}, 45, (1981) 389.

\bibitem{vWormer94}59. L. van Wormer, J. Goerres, C. Iliadis, M. Wiescher \&
F. K. Thielemann, {\it Ap. J.} 432, (1994) 326.

\bibitem{Wei37}60. C. F. von Weizs\"acker, {\it Phys. Z.},
38, (1937) 176.

\bibitem{Wei38}61. C. F. von Weizs\"acker, {\it Phys. Z.} 39, (1938) 633.

\bibitem{Woo86}62. S. E. Woosley and T. A. Weaver, {\it Ann. Rev. Astr. Astrophys.},24 (1986) 205.

\bibitem{Cha92}63. A. E. Champagne and M. Wiescher, {\it Ann Rev. Nucl. Part. Sci} 42, (1992) 39.

\end{thebibliography}
\end{document}